\documentclass{aa}  

\usepackage{natbib}
\bibpunct{(}{)}{;}{a}{}{,}

\usepackage{graphicx}

\usepackage{natbib,twoopt}
\usepackage[breaklinks]{hyperref}      

\usepackage{soul}
\usepackage{orcidlink}

\definecolor{cobalt}{rgb}{0.06, 0.2, 0.65}
\hypersetup{
  colorlinks,
  citecolor=cobalt,
  linkcolor=[rgb]{0.8, 0.2, 1.0},
  urlcolor=cobalt,
}

\usepackage[labelfont=bf]{caption} 
\usepackage[varg]{txfonts}

\usepackage{tikz}
\usetikzlibrary{shapes.geometric, arrows}
\tikzstyle{startstop} = [rectangle, rounded corners, minimum width=1cm, minimum height=1cm,text centered, text width=1cm, draw=black, fill=red!20]
\tikzstyle{io} = [trapezium, trapezium left angle=70, trapezium right angle=110, minimum width=3cm, minimum height=1cm, text centered, text width=3cm, draw=black, fill=blue!10]
\tikzstyle{process} = [rectangle, minimum width=3cm, minimum height=1cm,text centered, text width=3cm, draw=black, fill=orange!20]
\tikzstyle{process2} = [rectangle, minimum width=3cm, minimum height=1cm, text centered, text width=3cm, draw=black, fill=orange!20]
\tikzstyle{decision} = [diamond, minimum width=2cm, minimum height=1cm, text centered, text width=2.5cm, draw=black, fill=purple!10]
\tikzstyle{arrow} = [thick,->,>=stealth]

\usepackage{amsmath}
\usepackage[font=small,skip=3pt]{caption}
\usepackage[switch]{lineno}

\begin{document}

\title{Testing masking effectiveness using multi-line image cubes based on COSMOS2020 for [CII] line intensity mapping at $z_{[\textrm{CII}]}>3.5$}

\author{J. Clarke\inst{1}\thanks{Corresponding author: \email{jclarke@ph1.uni-koeln.de}}\orcidlink{0009-0006-6570-5804} \and C. Karoumpis\inst{2}\orcidlink{0000-0003-3259-7457} \and A. Dev\inst{2}\orcidlink{0009-0008-6563-3681} \and D. Riechers\inst{1}\orcidlink{0000-0001-9585-1462} \and T. Oak\inst{2}\orcidlink{0009-0001-9456-1401} \and Y. Okada\inst{1}\orcidlink{0000-0002-6838-6435} \and K. Narita\inst{3,4}\orcidlink{0009-0000-5913-8555}\and F. Bertoldi\inst{2}\orcidlink{0000-0002-1707-1775}}
\institute{I. Physikalisches Institut, Universität zu Köln, Zülpicher Straße 77, D-50937 Köln, Germany\and Argelander-Institut für Astronomie, Universität Bonn, Auf dem Hügel 71, 53121 Bonn, Germany\and Department of Astronomy, Graduate School of Science, The University of Tokyo, 7-3-1 Hongo, Bunkyo-ku, Tokyo 133-0033, Japan\and National Astronomical Observatory of Japan, National Institutes of Natural Sciences, 2-21-1 Osawa, Mitaka, Tokyo 181-8588, Japan}
 
\abstract 
{}
{We created line intensity mapping (LIM) intensity cubes for CO and [CII] emission lines using the COSMOS2020 galaxy catalogue, forming predictions based on empirical data, for observations from Prime-Cam mounted on the Fred Young Submillimetre Telescope (FYST). We also included simulated noise including white and correlated components, and tested masking techniques to recover [CII] signal at $3.5<z<8.2$.}
{We applied line luminosity models to the COSMOS2020 galaxy catalogue, spanning $1.44\,\textrm{deg}^2$ on-sky, to estimate the [CII] and CO $J$\,=\,1$-$0 emission, with other CO transitions derived using Spectral Line Energy Distribution (SLED) templates. From these we made cubes for four $\sim$40\,GHz bands in the EoR-Spec frequency range ($205-420$\,GHz), as well as a potential future upgrade to EoR-Spec including bands at 150 and 90\,GHz. Given the incompleteness of the empirical catalogue, these predictions are conservative lower limits, which we subsequently extrapolated from. We applied masks to recover the [CII] power spectra, using bright galaxies of COSMOS2020 as a foreground catalogue to target CO at low $z$ (targeted masking), and matching bright voxels across frequency bands to eliminate those associated with CO emission (blind masking).}
{Our CO intensity cube predictions are consistent with ALMA, VLA and NOEMA observations at high spatial resolutions,
indicating that this method gives realistic CO estimates within E-COSMOS. In ideal conditions, masking can recover [CII] above 300\,GHz for most models, with targeted masking requiring a complete foreground catalogue of bright CO sources to prevent them from contributing to contaminant emission, and blind masking needed additional lower frequency ranges to be effective. However, the combined noise will hinder [CII] recovery above 300\,GHz until near the end of the currently planned 2000 hour observing period, as $S/N<5$ without additional observing time. Whilst CO can be recovered below 300\,GHz, [CII] will likely be unavailable without the use of cross correlation techniques.}
{}


\titlerunning{Masking effectiveness when applied to [CII] and CO image cubes}
\authorrunning{J. Clarke et al.}

\maketitle
%
\section{Introduction} \label{sec:introduction} 

Line intensity mapping (LIM) can determine information from galaxies even without resolving them, providing a distinct niche from traditional surveys \citep{Bernal_2022}. LIM observations scan fields of several $\textrm{deg}^2$ using a wide beam and large frequency bins. By combining angular and spectral information, LIM probes three-dimensional cosmological volumes on Mpc scales, and so recovers the aggregate intensity from galaxy emission integrated over the redshift and sky angle covered by a resolution element. The three-dimensional (3D) spherically averaged power spectra over the resulting line emission intensity cubes determines intensity variance on different spatial scales, each component delineating contributions from bright and dim sources. In this way LIM can constrain the line luminosity function of a field whilst scanning faster than conventional surveys. 

Various frequency regimes can be used to probe different eras of the universe, with submillimeter (submm) being suitable in investigating the Epoch of Reionisation (EoR, \citealt{Zaroubi_2013}). This is when the neutral medium post-recombination was ionised by high-energy photons emitted from early galaxies at redshifts $z>6$, so line emission traces early galaxy formation and ionisation history of the intergalactic medium before cosmic noon ($z\approx2$, \citealt{Madau_2014}). LIM is uniquely suited to this considering how its wide area scans minimize sample variance, which high-resolution instruments such the Atacama Large Millimeter/submillimeter Array (ALMA) and \textit{James Webb} Space Telescope (JWST) cannot match in a timely fashion.

When observing LIM in the submm regime, one of the most important emission lines is the ionised carbon fine structure line ([CII], $\nu_{\textrm{rest}}=1900.537$\,GHz) which traces the star formation rate (SFR) of galaxies. [CII] is particularly difficult to recover at high $z$ with conventional surveys \citep{Schouws_2023}, making it a key target of multiple LIM experiments. These include CarbON [CII] line in post-rEionization and ReionizaTiOn epoch (CONCERTO, \citealt{CONCERTO_Collaboration_2020}), Tomographic Ionized-Carbon Mapping Experiment (TIME, \citealt{Crites_2014}), EoR Deep Spectroscopic Survey (EoR-Spec DSS), Terahertz Intensity Mapper (TIM, \citealt{Vieira_2020}), and Terahertz Integral Field Unit with Uni-
versal Nanotechnology (TIFUUN, \citealt{Rybak_2024}). The EoR-Spec DSS utilises the EoR-Spec module of the Prime-Cam instrument (\citealt{Choi_2020}, \citealt{Nikola_2023}, \citealt{Freundt_2024}) mounted on the ground-based Fred Young Submmillimeter Telescope (FYST), a key part of the CCAT project (\citealt{CCAT_Prime_Collaboration_2022}). It will probe [CII] signal from $210-420$\,GHz ($3.5<z_{\textrm{[CII]}}<8.2$), at the end of the EoR, over the two $2\times2=4\textrm{\,deg}^2$ fields E-COSMOS and E-CDF-S (Section \ref{sec:methodSampleBase}), starting in 2027. These fields also have ancillary data from traditional galaxy surveys (COSMOS: \citealt{Scoville_2007}, \citealt{Weaver_2022, Weaver_2023}, \citealt{Shuntov_2025}; Euclid: \citealt{Euclid_2022}, \citealt{Euclid_2025}, \citealt{Euclid_2025_main}). Other lines observed within submm LIM include the carbon monoxide rotational transitions (CO, $\nu_{\textrm{rest}}=115.27$\,GHz for the J=1-0 transition), explored by experiments such as CO Mapping Array Pathfinder (COMAP, \citealt{Cleary_2022}), as well as the doubly ionised oxygen line ([OIII], $\nu_{\textrm{rest}}=3393$\,GHz).

Simulations predicting the target signal (e.g. \citealt{Chung_2020}, \citealt{Karoumpis_2021, Karoumpis_2024}, \citealt{Bethermin_2022}, \citealt{Roy_2023}, \citealt{VanCuyck_2023}, \citealt{Carlson_2025}) are being made to prepare for these LIM experiments. Prior simulations used hydrodynamical simulation tools such as EAGLE (\citealt{EAGLE_2017}) and IllustrisTNG (\citealt{Nelson_2019}) to generate dark matter haloes and populate them with galaxies, which they paint line luminosities onto using empirical models derived from observations. There are also fully analytic treatments starting from the underlying cosmological assumptions, deriving the power spectra and intensity cubes using codes such as \texttt{lim} and \texttt{oLIMpus} (\citealt{Bernal_2019}, \citealt{Libanore_2025}). 

\cite{Clarke_2024}, from now on C24, provided an empirical grounding for the simulation-based estimates by using existing galaxy catalogue data for their simulations. Utilising galaxy location data and photometric measures of properties such as stellar mass and star formation rate (SFR), they applied line luminosity models (such as \citealt{Silva_2015} and \citealt{Romano_2022}) to paint line luminosity onto mock intensity cubes. This methodology produces lower limits for power spectra as the catalogues only include galaxies from conventional observations, and so are inherently incomplete. To address this, C24 then extrapolated from the Cosmic Evolution Survey 2020 galaxy catalogue (COSMOS2020, \citealt{Weaver_2022}) using the Cosmic Assembly Near-infrared Deep Extragalactic Legacy Survey (CANDELS, \citealt{Nayyeri_2017}), thereby estimating the contribution from undetected galaxies and producing [CII] predictions for EoR-Spec in the E-COSMOS field which were consistent with prior work. While this proof-of-concept predicted that [CII] detections by FYST would be feasible for the post-EoR era ($>$300\,GHz, z$<$5), it did not include any contaminants.

It is vital to include contaminants in simulations, and to develop techniques to remove them, because LIM observations cannot inherently distinguish them from the target signal. They includes large scale foregrounds such as the Milky Way interstellar medium, other line emission from different redshifts, as well as instrumental red noise, atmospheric noise, scan pattern and edge effects in observations. When other sources of line emission are the primary contaminant, masking techniques can recover the target signal by excluding specific voxels which we know are dominated by their signal. In prior work, masking techniques which remove bright voxels or known contaminants via a flux cut were effective (\citealt{Gong_2014}, \citealt{Silva_2018}, \citealt{Gkogkou_2023}), but this runs the risk of eliminating bright [CII]. Another approach uses a catalogue of foreground galaxies (``foreground catalogue'') to target and remove contaminant sources from low $z$ galaxies via ``targeted masking'' (\citealt{Yue_2015}, \citealt{VanCuyck_2023} and \citealt{Karoumpis_2024}). Regardless of methodology, simulating contaminant removal is key to optimise LIM experiments prior to observations, and it can mitigate ``unknown unknowns'' found by pathfinder experiments. 

In this work we extend the methodology of C24 by including CO, a prominent source of contaminant in the foreground (primarily $z<2$) within the frequency range of EoR-Spec. We use COSMOS2020 as a known foreground catalogue to implement targeted masking techniques on these intensity cubes, to test recovering the [CII] power spectra. As COSMOS2020 is inherently incomplete, we also use it to investigate the impact of incompleteness of foreground catalogues for masking. We include two lower frequency bands as a hypothetical extension to EoR-Spec, as done by \cite{Roy_2024}, which enables other signal cleaning techniques such as blind masking. In addition, we incorporate early estimates of EoR-Spec white and correlated noise (from Dev et al. in prep.) and analyse the challenges it poses for detecting power spectra. In this way we show the viability of this method when applied to wide-field surveys and multiple emission lines, with the possibility of investigating future cross-instrument cross-correlation.

Section \ref{sec:method} describes creating the intensity cubes for EoR-Spec observations from COSMOS2020, by taking the appropriate galaxies and applying bulk property models. This includes the galaxy extrapolation procedure and CO signal. It also describes the masking formulation, and the procedure of applying these masks. Section \ref{sec:results} analyses the CO signal in the cubes, including its relative strength to [CII], before and after applying targeted or blind masks, for empirically-based lower limits and empirically-anchored extrapolated samples. We also include comparisons to white and correlated noise. Then, Section \ref{sec:discussion} discusses the challenges from our analysis: working with a small field, having an incomplete foreground catalogue, and mitigating the impact of white and correlated noise.

We use a flat $\Lambda$CDM dark matter universe, taking $H_0=70 \textrm{\,kms}^{-1}\textrm{Mpc}^{-1}$, $\Omega_\textrm{M}=0.3$, and $\Omega_\Lambda=0.7$, though we did not find meaningful differences when using other cosmologies (e.g. \citealt{Planck_2020}). We use magnitudes in AB format \citep{Oke_1983}. In this work all power spectra are ``auto-spectra'', where we correlate a map with itself, as opposed to the ``cross-spectra'' between two different maps.
\section{Method} \label{sec:method}

\begin{figure*}[t]
\centering
\begin{tikzpicture}[node distance=1.75cm]

\node (start) [startstop] {Start};
\node (checkedmadetomogs) [decision,below of=start,yshift=-1.4cm] {Have combined intensity cubes for all frequency bands been made?};
\node (checkedmadelines) [decision,right of=checkedmadetomogs,xshift=3.2cm] {Have cubes for all lines in the band been made?};
\node (takecatalogues) [io,right of=checkedmadelines,xshift=3cm] {Take appropriate catalogues for the redshift (Secs. \ref{sec:methodSampleBase}, \ref{sec:methodSampleExtrap})};
\node (takeapplymodel) [process,right of=takecatalogues,xshift=2.55cm] {Apply line model to catalogue \\to form intensity cube (Sec. \ref{sec:methodTomog})};
\node (formmasks?) [decision,below of=takeapplymodel,yshift=-2.49cm] {Is it a CO transition?};
\node (formmasksyes) [io,left of=formmasks?,xshift=-2.55cm] {Form targeted mask (Sec. \ref{sec:methodMasksTarget})};
\node (inclexclwn?) [decision,below of=checkedmadelines,yshift=-2.49cm] {Do we include atmospheric noise?};
\node (inclwnyes) [io,below of=inclexclwn?,yshift=-1.15cm] {Take noise parameters, and make noise cube (Sec. \ref{sec:methodNOISE})};
\node (formtom) [process,left of=inclwnyes,xshift=-3.2cm] {Sum all cubes to make cube (Sec. \ref{sec:methodTomog})};

\node (targetedmasking) [process,above of=formtom,yshift=1.09cm] {Analyse cubes (Sec. \ref{sec:results_CO}), perform targeted masking technique (Fig. \ref{fig:FigFlowchart_TargetedMasking})};

\node (blindmask) [process,above of=checkedmadetomogs,yshift=1.4cm,xshift=4.9cm] {Blind Masking Technique (Fig. \ref{fig:FigFlowchart_Blindmasking})};
\node (stop) [startstop,xshift=3cm,right of=blindmask] {End};

\draw [arrow] (start) -- (checkedmadetomogs);
\draw [arrow] (checkedmadetomogs) --  node[anchor=south] {no} (checkedmadelines);
\draw [arrow] (checkedmadelines) --  node[anchor=south] {no} (takecatalogues);
\draw [arrow] (takecatalogues) -- (takeapplymodel);
\draw [arrow] (takeapplymodel) -- (formmasks?);
\draw [arrow] (formmasks?) -- node[anchor=north] {yes} (formmasksyes);
\draw [arrow] (formmasks?) -- node[anchor=south] {no} (checkedmadelines);
\draw [arrow] (formmasksyes) -- (checkedmadelines);
\draw [arrow] (checkedmadelines) -- node[anchor=west] {yes} (inclexclwn?) ;
\draw [arrow] (inclexclwn?) -- node[anchor=west] {yes} (inclwnyes) ;
\draw [arrow] (inclexclwn?) -- node[anchor=west] {no} (formtom)  ;
\draw [arrow] (inclwnyes) -- (formtom) ;
\draw [arrow] (formtom) -- (targetedmasking) ;
\draw [arrow] (targetedmasking) -- (checkedmadetomogs) ;
\draw [arrow] (checkedmadetomogs) -- node[anchor=west] {yes} (blindmask) ;
\draw [arrow] (blindmask) -- (stop) ;
\end{tikzpicture}
\caption{Flowchart showing the steps of forming our combined intensity cubes and masks.}
\label{fig:FigFlowchart_Main}
\end{figure*}
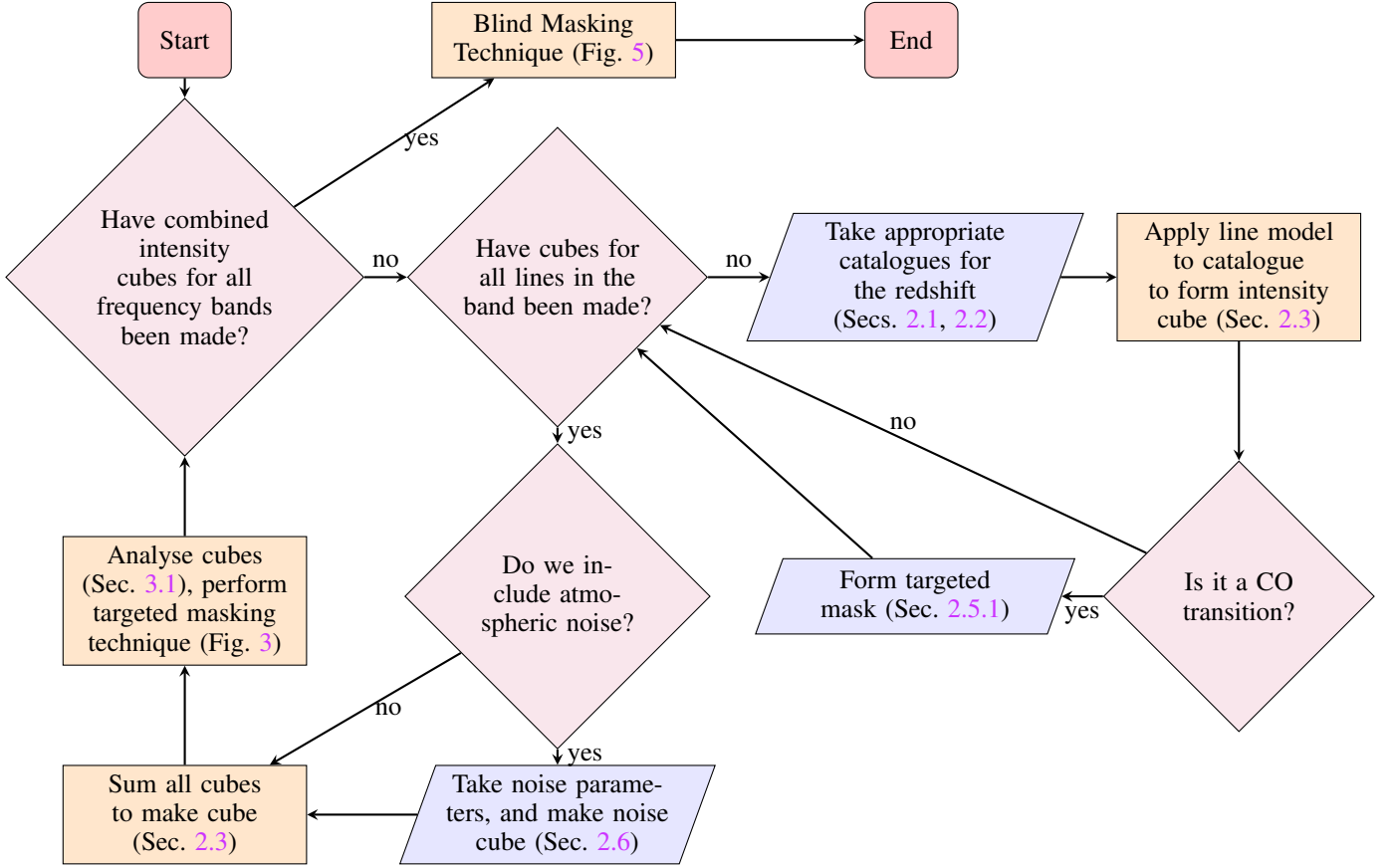

Our procedure is summarised in Fig. \ref{fig:FigFlowchart_Main}. For each frequency band, we create intensity cubes for all relevant emission lines, and corresponding noise cubes. We take catalogue data from the redshifts relevant to each line, and form an intensity cube by applying an appropriate line model. Using the same galaxy data we also form targeted masks for that line. We then combine these cubes, including noise when relevant, before analysing them. As part of our analysis we also apply the targeted masking technique as described in Fig. \ref{fig:FigFlowchart_TargetedMasking}. Once we have made complete intensity cubes for all frequency bands, we also match their voxels to perform blind masking, described in Fig. \ref{fig:FigFlowchart_Blindmasking}.

\subsection{Base sample}\label{sec:methodSampleBase}

We used the COSMOS2020 galaxy catalogue (version 4.1.1\footnote{https://irsa.ipac.caltech.edu/data/COSMOS/overview.html}) to create intensity cubes for the E-COSMOS field, following C24. The sample contains 1.7\,million galaxies in a $1.7\times2\textrm{\,deg}^2$ field ($149-151\textdegree$ RA, $1.4-3.1\textdegree$ Dec) for $z<6.3$, with four UltraDeep stripes of Ultra-VISTA covering $0.7\textrm{\,deg}^2$ for $6.3<z<9$, with cutouts to remove noise from foreground stars \citep{Weaver_2022}. As with C24, we used the smaller UVISTA area of $1.2\times1.2=1.44\textrm{\,deg}^2$ ($149.6-150.8\textdegree$ RA, $1.6-1.8\textdegree$ Dec) to remove edge artefacts, and exclude all galaxies with IRAC Channel 1 AB mags above 26. We took galaxy bulk properties from THE FARMER, using photometry from \texttt{The Tractor} code \citep{Lang_2016} to follow the completeness statistics of \cite{Weaver_2023} and methods of C24. The primary photometric properties for each galaxy are calculated using the LePhare photometric code \citep{Arnouts_2002}, with redshift, SFR, stellar mass, and IRAC Channel 1 AB mags. However, we took $L_{\textrm{[OIII]}}$ and $L_{\textrm{IR}}$ $8-1000\,\mu$m luminosity measurements from the EAZY code \citep{Brammer_2008} as they are not included in THE FARMER. We refer to this sample as FARMER LP.
\subsection{Extrapolated samples}\label{sec:methodSampleExtrap}

As FARMER LP is a traditional galaxy survey, it is intrinsically incomplete and therefore misses faint and dust-obscured galaxy populations which are not bright in UV. In this way any predictions made from it are ``lower limits'', a strict empirical lower bound on the line intensity that we expect (for a given line model). This is useful for considering the most pessimistic case for recovering signal, however its incompleteness means it is unlikely to be representative of that patch of sky. Extrapolation attempts to correct for this, adding additional galaxies at the faint and bright ends of the luminosity function, thereby producing samples we expect to be more representative of reality. While the process of extrapolation described below is phenomenological and strongly dependent on the exact methodology used, especially below the mass completeness limit, it being anchored to both FARMER LP and deeper samples within the COSMOS field gives us confidence in our methods. This is especially true considering that C24 demonstrated that extrapolation from COSMOS2020 produces predictions compatible with conventional simulations. Most results in Section \ref{sec:results} are therefore based on the extrapolated sample, with the lower limits mainly used for comparison.

The method described in this work is similar to that of C24, that is duplication of existing galaxies within FARMER LP until the sample abides by completeness statistics however we had to modify this to account for multiple emission lines. Notably, we did this for the entire sample in $\Delta z=0.5$ redshift intervals from $z=0$ to $z=9$, instead of just the four [CII] redshift ranges in the EoR-Spec frequency range. The full process is described below.

We first included galaxies at the positions of masked or unobserved regions in FARMER LP, primarily due to foreground stars. We achieved this by randomly duplicating galaxies from the sample to maintain the same number density, and placing them at random positions within the masked regions. For $z<6.3$, this area is $\sim$10\% of the on-sky map, but for $z>6.3$ we also accounted for the $\sim$45\% area lost due to only covering the UltraDeep stripes. The original sample combined with stellar mask extrapolation is treated as the ``base sample''.

Then we account for the completeness of FARMER LP above and below the mass completeness limit, the $\sim$70\% completeness mass limit described by \citep{Weaver_2023}:
\begin{equation}
\frac{M_{\textrm{lim}, \star}}{M_\odot}= -3.23 \times 10^7 (1 + z) + 7.83 \times 10^7 (1 + z)^2, 
\label{eq:WeaverComplete}
\end{equation}

To be able to use this $70\%$ completeness ratio to apply a correction to galaxy numbers above the mass completeness ratio, we need to verify it for our specific subsample. This was achieved by calculating the number ratio of galaxies above this mass limit between FARMER LP and CANDELS (Table \ref{table:CANDELSlims}, \citealt{Nayyeri_2017}). As CANDELS is a deep field survey covering a $0.06\textrm{\,deg}^2$ region within COSMOS, with an higher average magnitude limit (up to $28$\,mag compared to $26$\,mag), 
for the purpose of this paper we use it as point of comparison which is fully complete.
From our investigation we found the mass completeness of FARMER LP approaches $50\%$ for $z<1$ and $z>5.5$, with $2<z<3$ approaching $100\%$. Note that for $z>6.5$ we assume a fixed $50\%$ mass completeness following $5.5<z<6.5$, as both FARMER LP and CANDELS trace the same few bright galaxies, and CANDELS cannot be treated as $100\%$ complete any more. From these ratios we determined the number of galaxies we need: for example, with a completeness of 0.72, we generated $1/0.72-1\approx0.4$ additional galaxies for each existing galaxy above the mass limit in FARMER LP. As high mass galaxies are typically more luminous and contribute more towards the power spectra \citep{VanCuyck_2023}, we expect lines at $z<1$ and $z>5.5$ to show a greater increase in signal when extrapolated compared to other lines. We note these ratios follow the star formation rate density plot of \cite{Madau_2014}, indicating that FARMER LP is more complete around cosmic noon.

\begingroup
\setlength{\tabcolsep}{3pt} 
\begin{table*}
\caption{For each $\Delta z=0.5$ interval: 70\% stellar mass completeness COSMOS2020 \citep{Weaver_2022, Weaver_2023}, ratios of FARMER LP to CANDELS galaxies \citep{Nayyeri_2017} above the completeness limits, galaxy main sequence parameters, and Schechter fit parameters to the mass functions of each interval. For $z>3.5$ the Schechter fit parameters are designed to fit the low mass end of data points from COSMOS2025 \citep{Weibel_2024, Shuntov_2025}, hence the fit parameters. For $z\geq6.5$ the logarithmic mass limit increases by 0.1 for each $\Delta z=0.5$.}
\centering
\begin{tabular}{c | c c | c c  | c c c}
\hline\hline
 $z$ & $\log \frac{M_\star}{M_\odot}$ & CANDELS Ratio & $\alpha$ & $\beta$ & $\log \frac{\Phi_0}{M_c/M_\odot}$ & $\alpha$ & $\log \frac{M_c}{M_\odot}$ \\
\hline
$0\leq z<0.5$ &7.91* &0.49 &-6.046&  0.5887 &-2.16& 0.343& 11.06\\
$0.5\leq z<1$ &8.26 &0.57 &-4.866&  0.5322 &-2.13&  0.267& 11.06\\
$1\leq z<1.5$ &8.51 &0.82 &-4.913&  0.5768 &-2.59&  0.347& 11.16\\
$1.5\leq z<2$ &8.70 &0.72 &-5.425&  0.6509 &-2.48&  0.106& 11.00\\
$2\leq z<2.5$ &8.85 &0.94 &-5.547&  0.6768 &-3.10&  0.444& 11.15\\
$2.5\leq z<3$ &8.99 &1.00 &-5.707&  0.7082 &-3.39&  0.580&11.20\\
$3\leq z<3.5$ &9.11 &0.70 &-5.628&  0.7020 &-3.51& 0.568& 11.36\\
$3.5\leq z<4$ &9.21 &0.82 &-5.209&  0.6657 &-4.58&  0.838& 11.38  \\
$4\leq z<4.5$ &9.30 &0.87 &-4.861&  0.6410 &-4.58&  0.838& 11.38  \\
$4.5\leq z<5$ &9.38 &0.78 &-4.527&  0.6093 &-16.6 &  1.21 & 20.53\\
$5\leq z<5.5$ &9.46 &0.73 &-3.837&  0.5322 &-16.6 &  1.21 & 20.53\\
$5.5\leq z<6$ &9.52 &0.45 &-4.581&  0.6057 &-32.5&   1.08&  36.31\\
$6\leq z<6.5$ &9.59 &0.48 &-4.951&  0.6525 &-32.5&   1.08&  36.31\\
$z\geq6.5$ &9.6+(0.1$\times(z-0.625)$) & 0.5**  &-4.951&  0.6525 &-14.8&   0.979&  20.93\\
\hline
\end{tabular}
\tablefoot{*See Section \ref{sec:methodMasksTarget} for implementation in masking for $z<0.5$. **In actuality this is $\sim$1, fixed to 0.5 as discussed in main body.}
\label{table:CANDELSlims}
\end{table*}
\endgroup

We also modelled the galaxies below the mass completeness limits. C24 fitted Schechter functions to the [CII] line luminosity function, however we made this line-independent by instead fitting to the base+CANDELS galaxy stellar mass function. The Schechter function is defined as follows \citep{Schechter_1976}:
\begin{align}
\Phi(M)\textrm{d}M&=\left(\frac{\Phi_0}{M_\textrm{c}/M_\odot}\right) \left(\frac{M}{M_\textrm{c}}\right)^\alpha e^{-\frac{M}{M_\textrm{c}}}\textrm{d}M,
\label{eq:SchechterFit}
\end{align}
\noindent where $\Phi$ is the number of galaxies per unit comoving volume in $\textrm{Mpc}^3$ per unit dex, $\Phi_0$ and $\alpha$ are scaling factors, and $M_c$ determines the ``knee'' between the two regimes of the Schechter curve. These fittings are simple below $z<3.5$ due to the clearly defined knees (see Table \ref{table:CANDELSlims}) but at higher redshift the fits are less reliable due to having less information below the knee. Instead, we fitted to mass function parameters for COSMOS galaxies taken from JWST for $3.5<z<9.5$, COSMOS2025 (given in Tables 2 and 3 by \citealt{Weibel_2024}). As the aim here was to appropriately fit below the mass completeness limits $\sim$$10^9\,M_\odot$, the solutions do not need to be plausible for high masses.

Subsequently we added galaxies to the sample to abide by the Schechter fit, for each mass bin down to $10^8\,M_\odot$. Modelling galaxies with lower masses was unsuitable as our understanding of emission of dwarf galaxies at high redshifts is limited to serendipitous findings (e.g. \citealt{Gelli_2020}). In addition, models from the literature which do attempt to model them find their contribution to be negligible \citep{VanCuyck_2023}. 
From empirical testing, we found that bulk properties of galaxies below the mass completeness limits have higher uncertainties with more outliers, especially for SFR. As in C24, we performed best fits between stellar mass and SFR for galaxies in FARMER LP in each $\Delta z=0.5$ interval (Eq. \ref{eq:GalMainSequence}, with $a$ and $b$ in Table \ref{table:CANDELSlims}). From the stellar masses of the newly extrapolated galaxies we assigned SFRs accordingly, adding 0.5\,dex scatter to ensure variance consistent with the existing relationship. 
\begin{align}
\log_{10}\frac{\textrm{SFR}}{M_\odot \textrm{yr}^{-1}}=\alpha+\beta\log_{10} \frac{M_\star}{M_\odot},
\label{eq:GalMainSequence}
\end{align}

We determined the locations of extrapolated galaxies by following the Voronoi Tessellation method as described in C24, using different distribution maps for each $\Delta z=0.5$ interval. To summarise, a 3D cube based on the stellar mass of base sample galaxies weighted the random distribution of the new galaxies, mimicking how galaxies cluster in dark matter haloes. When these additional extrapolated galaxies are combined with the base, it is henceforth referred to as the extrapolated sample.
\subsection{Constructing multi-line intensity cubes}\label{sec:methodTomog}

Our procedure of creating an intensity cube from a galaxy catalogue is similar to C24, with adjustments made to facilitate masking. In that work they took galaxies at redshifts where the observed [CII] emission aligns with the relevant frequency band (Eq. \ref{eq:FreqZRelation}). They calculated the [CII] luminosity for each galaxy using a given bulk property model and then found the intensity. These intensities were inserted into a voxel of a 3D cube representing the map over the observed band, where each voxel corresponds to the area covered by the full-width half maximum (FWHM) of the on-sky beam in angular space and frequency channel in frequency space. The voxel chosen corresponds to the observed frequency and on-sky position of a galaxy.

We adapted this method to create a multi-line intensity cube, incorporating CO transitions with [CII]. The frequency space covered by our six bands is $390-418$, $330-370$, $260-300$, $205-245$, $130-170$, and $70-110$\,GHz, with the first four corresponding to bands used in the literature within the continuous spectral coverage of EoR-Spec (\citealt{Choi_2020}, \citealt{CCAT_Prime_Collaboration_2022}). They were primarily chosen to avoid poor atmospheric transmission, though depending on specific conditions at the site the accessible range may be widened or shrunk. 

The 90 and 150\,GHz bands will not be initially accessible by EoR-Spec, but were included as they have already been explored as an extension for Prime-Cam \citep{Roy_2024}. As well as having excellent atmospheric transmission, their lower frequencies observe CO transitions at higher redshift (e.g. $1.7<z_\textrm{CO(4-3)}<5.5$ compared to $0<z<2.2$), providing more than a dozen new cross-correlation power spectra. While we do not explore this in our work, our blind masking technique works using similar principles, and adding these bands improves contaminant cleaning by up to 0.5\,dex (Section \ref{sec:results_blindmasking}). Outside of Prime-Cam, those bands partially overlap with the frequency range covered by TIFUUN, so our work can be considered a first-order estimate of results for their instrument \citep{Rybak_2024}. 

\begin{figure}[t]
 \centering
 \includegraphics[width=\linewidth]{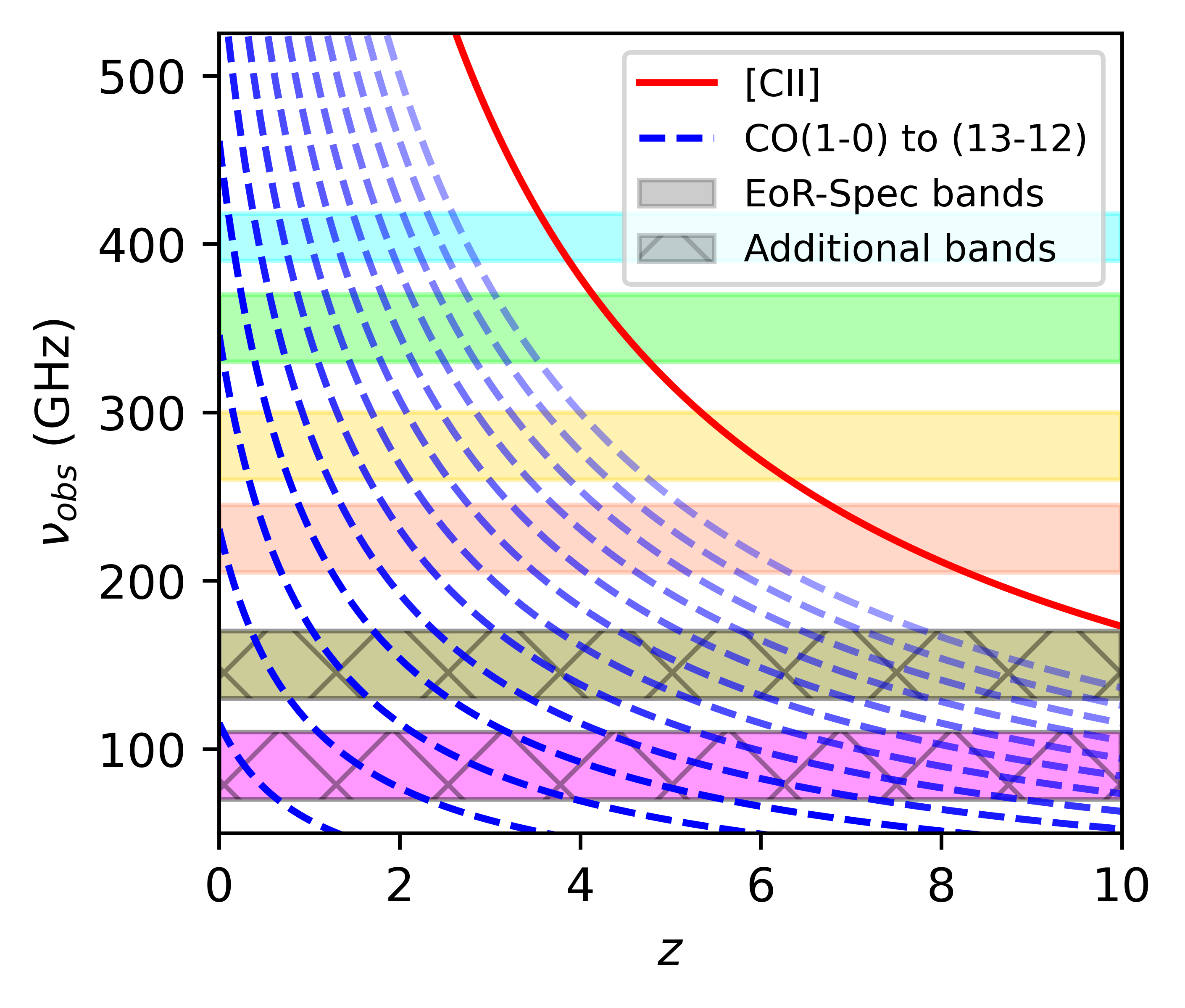}
\captionof{figure}{Change of observed frequency with redshift for [CII] and the CO rotational transition lines up to $J$\,=\,13$-$12. The shaded horizontal bars show the four bands nominally covered by EoR-Spec (\citealt{CCAT_Prime_Collaboration_2022}), with the hatched ones covering the extension from \cite{Roy_2024}. The lines show how redshift changes with observed frequency for the emission lines (CO blue/dashed, [CII] red/solid).}
 \label{fig:ZvsFreq}
\end{figure}

For each frequency range, we created separate intensity cubes for each individual line, applying the appropriate bulk-property models to galaxies with redshifts corresponding to the observed frequency range and the line rest frequency as below: 
\begin{equation}
z=\frac{\nu_{\textrm{rest}}}{\nu_{\textrm{obs}}}-1,
\label{eq:FreqZRelation}
\end{equation}

\noindent where $\nu_\textrm{rest}$ is the line rest frequency, and $\nu_\textrm{obs}$ is the observing frequency. Fig. \ref{fig:ZvsFreq} shows the redshift range of galaxies for each line in each frequency band.

\begingroup
\setlength{\tabcolsep}{3pt} 
\begin{table*}[t]
\caption{Intensity cube properties, dimensions and assumed white noise power spectra for each EoR-Spec frequency band and its extensions.}
\centering
\begin{tabular}{c c c c c c c}
\hline\hline
 Freq Bands (GHz) & $390-418$ (1) & $330-370$ (1)& $260-300$ (1)& $205-245$ (1)& $130-170$ (2)& $70-110$ (2)\\
\hline
Beam FWHM (arcsec) & 33 & 37 & 48 & 58 & 69 & 78 \\
Freq. Bin Width (GHz) & 4.1 & 3.5 & 2.8 & 2.2 & 1.5 & 0.9 \\
Cube Dim. (w/ subgrid) &$21\times396\times396$ & $33\times351\times351$& $42\times273\times273$& $54\times225\times225$& $81\times189\times189$& $132\times168\times168$ \\
$P_\textrm{N}$ (Mpc$^3$ Jy$^2$ sr$^{-2}$) & $1.2\times10^{11}$ & $3.9\times10^{10}$ & $4.9\times10^{9}$ & $2.6\times10^{9}$ & $1.1\times10^{9}$ & $9.2\times10^{8}$ \\
\hline
\end{tabular}
\tablebib{(1) \cite{CCAT_Prime_Collaboration_2022}, (2) \cite{Roy_2024}}
\label{table:FreqData}
\end{table*}
\endgroup

The bulk property emission models are not derived from the small-scale processes which cause the specific emission, and so are largely unconstrained. In this context using a wide range of models for each line is the best way to provide reasonable upper and lower constraints. To explore the parameter space, we used three models for the [CII] and CO lines, covering bright, faint, and intermediate cases. These models were all verified: the [CII] models were shown to align with the literature in C24, and we compare the CO line luminosity functions from our models to observations from ALMA, the VLA and NOEMA in Section \ref{sec:results_CO}. 

The lower limit [CII] model we used was a SFR-$L_\textrm{[CII]}$ power law from \cite{DeLooze_2014} using its entire sample of 530 dwarf galaxies, including 27 at $z>0.5$ (``DL14 Entire''). This type of model is justified from low redshift observations \citep{Agrawal_2025}. 
We use this specific model instead of one based on the few high redshift galaxies, as applying the other model produces a brighter line luminosity function \citep{Yue_2019}, and we want DL14 Entire to produce a minimum limit. This selection of local [CII] galaxies contrasts with models based on ALPINE-ALMA galaxies at $4.4<z<5.9$ largely in COSMOS (\citealt{Bethermin_2020}, \citealt{Faisst_2020}, \citealt{Le_F_vre_2020}). Those include the ``m3'' model from C24, which forms an upper limit (``C24\_m3'') and was chosen due to having the highest luminosity function from that paper. It primarily depends on stellar mass, with a minor SFR dependency. Another ALPINE-ALMA model, based solely on SFR, is from \cite{Schaerer_2020} and forms a middle ground (``Sc20''). Each model is parametrised as a linear combination between $L_{\textrm{[CII]}}$, SFR and stellar mass, with coefficients $a \textrm{ and }b$ in Table \ref{table:CIImodel}, as below:
\begin{gather}
\log_{10} \frac{L_{[\textrm{CII}]}}{L_\odot}=a+b\log_{10}\frac{\textrm{SFR}}{M_\odot \textrm{yr}^{-1}}+c\log_{10}\frac{M_\star}{M_\odot},
\label{eq:CIIxSFRxMass}
\end{gather}

\begingroup
\setlength{\tabcolsep}{3pt} 
\begin{table}
\caption{Parameters for [CII] line emission bulk property models.}
\centering
\begin{tabular}{c c c c}
\hline\hline
 Model & $a$ & $b$ & $c$  \\
\hline
DL14 Entire (1) &6.99&1.01  &0\\
Sc20 (2) &6.43&1.26 &0\\
C24\_m3 (3) &3.30& -0.332& 0.601\\
\hline
\end{tabular}
\tablebib{(1) \cite{DeLooze_2014}, (2) \cite{Schaerer_2020}, (3) C24}
\label{table:CIImodel}
\end{table}
\endgroup

As Sc20 and C24\_m3 models were calibrated by galaxy data at $4.4<z<5.9$, they are therefore applicable to the EoR-Spec frequency ranges, though with greater uncertainty at 225 and 410\,GHz. DL14 Entire is less directly related, as it is based on smaller galaxies, but it forms a conservative lower limit.

When modelling CO, we implemented models for $L_{\textrm{CO\,} J=1-0}$, then propagated the other transitions' luminosities using a Spectral Line Energy Distribution (SLED) template, following the methodology of \cite{Karoumpis_2024}. They used a linear combination between the SLED of a bright galaxy (NGC253, derived from \citealt{Mashian_2015}) and a faint one (the Milky Way, derived by \citealt{Carilli_2013} and \citealt{Wilson_2017}), the exact combination determined by the location of a galaxy in the galaxy main sequence. This SLED parametrisation is defined fully in Appendix \ref{appendix: SLED}. The primary difference in our methodology is that we used bulk property models based on $L_{\textrm{IR}}$ for the starting $L_{\textrm{CO\,} J=1-0}$, as FARMER LP does not have accurate estimates of gas mass. We specifically use the two models from \cite{Sargent_2014} for main sequence and starburst galaxies as upper and lower limits (``SargMS'', ``SargSB''). SargMS, whilst split between galaxies at $z<0.1$ and $0.1<z<3.2$, forms a strong upper limit as other literature found their $L'_{\textrm{CO\,}J=1-0}/L_{\textrm{IR}}$ ratios to be four times higher than that of the starburst galaxies (\citealt{Genzel_2010}, \citealt{Prajapati_2025}). For a middle ground, we used a composite model derived by \cite{Li_2016} based on higher redshift galaxies from other literature (``Li''). These $L_{\textrm{IR}}$ models, with coefficients from Table \ref{table:COmodel}, follow:
\begin{gather}
\log_{10} \frac{L_{\textrm{CO\,}J=1-0}}{L_\odot}=d+e\log_{10}\frac{L_{\textrm{IR}}}{L_\odot},
\label{eq:COxLIR}
\end{gather}

\begingroup
\setlength{\tabcolsep}{3pt} 
\begin{table}
\caption{Parameters for CO $J$\,=\,1$-$0 bulk property models}
\centering
\begin{tabular}{c c c}
\hline\hline
 Model & $d$ & $e$  \\
\hline
SargSB (1) &-4.22&0.81 \\
SargMS (1) &-3.770&0.81 \\
Li (2) &-4.49&0.855 \\
\hline
\end{tabular}
\tablebib{(1) \cite{Sargent_2014}, (2) \cite{Li_2016}}
\label{table:COmodel}
\end{table}
\endgroup

$L_\textrm{CO}-L_\textrm{IR}$ power laws are a consequence of how the Kennicutt-Schmidt relation is observed over integrated intensity (\citealt{Schmidt_1959}, \citealt{Kennicutt_1998}), due to how gas surface density tightly relates to the underlying star formation processes. They are sensitive to the starting $L_{\textrm{IR}}$, so we treated outliers in FARMER LP carefully to ensure they did not skew the resulting power spectra. In particular we excluded galaxies below $z<0.1$ due to how the accuracy of bulk properties decreases (see Appendix \ref{appendix: low z}). For actual observations, it would be possible to mask these bright extended sources at low $z$ by traditional techniques, which we do not include here.

For the extrapolated galaxies, we used $L_{\textrm{IR}}$ values from fitting in line modelling, to avoid bias as discussed in Section \ref{sec:methodSampleExtrap}. These followed empirical SFR fits over mass-complete data from FARMER LP (Eq. \ref{eq:LIR-SFR}). We added 2\,dex worth of scatter, following fits derived using existing FARMER LP galaxies.
\begin{align}
\log_{10} \frac{L_{\textrm{IR}}}{L_\odot}=9.75+0.732\log_{10}\frac{\textrm{SFR}}{M_\odot \textrm{yr}^{-1}}, 
\label{eq:LIR-SFR}
\end{align}

After determining the luminosities of each galaxy, we converted them into intensities and inserted them into the 3D arrays representing the intensity cubes. The original voxel resolution for these arrays were determined by using one beam FWHM in angular map space and one unit of frequency resolution (Table \ref{table:FreqData}). However, to more effectively test masking we sub-grid these voxels by $3\times3\times3$ so that an individual voxel covers one-third of the beam FWHM and frequency bin, following \cite{Karoumpis_2024}. Using the given map position and redshift information, which we convert to observed frequency using Eq. \ref{eq:FreqZRelation}, we can determine the appropriate cube coordinates for each galaxy. Their intensities are given as:
\begin{gather}
\frac{I_{\textrm{Line}}}{L_\odot/\textrm{Mpc}^2/\textrm{GHz}/\textrm{rad}^2}=\frac{\textrm{rad}^2}{\Delta\theta_\textrm{subv.}^2}\frac{\textrm{GHz}}{\Delta\nu_\textrm{subv.}}\frac{L_{\textrm{Line}}}{L_\odot}\frac{\textrm{Mpc}^2}{4\pi r^2(1+z)^2},
\label{eq:IntensityGal}
\end{gather}

\noindent where $I$ is intensity, $\Delta\theta_\textrm{subv.}$ is the width of the map voxel (one-third of beam FWHM), $\Delta\nu_\textrm{subv.}$ is the smallest frequency unit (one-third of the frequency bin), $L_{[\textrm{Line}]}$ is the luminosity of the given line, and $r(1+z)$ is the luminosity distance with $r$ as comoving distance. We then converted these units to $\textrm{Jy\,sr}^{-1}$, our preferred intensity unit by convention. 

Following this, we convolved each voxel with the beam, a 2D Gaussian in map space and 1D Lorentzian in frequency space (\citealt{CCAT_Prime_Collaboration_2022}, Appendix \ref{appendix: Lorentzian V Gaussian}) for details. We included the Lorentzian spread in frequency space instead of a Gaussian as used in previous work (e.g. \citealt{Chung_2020}, \citealt{Karoumpis_2024}), as this is more representative of EoR-Specs spectral profile. Using this profile suppresses the 
power spectra by up to $\sim$$0.2\,\textrm{dex}$, as discussed in Appendix \ref{appendix: Lorentzian V Gaussian}, \cite{Marcuzzo_2025}, and Dev et al (in prep.).

Once an intensity cube was made for each line, we summed them to make the combined intensity cubes, resulting in [CII] cubes, summed CO cubes, and total cubes. Note that beam convolution before or after summing the individual components has no impact. Some intensity cubes also include white and correlated noise, discussed in Section \ref{sec:methodNOISE}. We did not include additional effects due to the calibration and data reduction process, and used idealised instrument parameters. 
\subsection{Power spectra}\label{sec:methodPS}

The 3D spherically averaged power spectra statistic $P(k)$ investigates the intensity cubes by probing the galaxy line luminosity function and the scale of galaxy clustering. Because of how the power spectra is weighted by the luminosity of individual sources, the first and second moment of the luminosity function inform the magnitude of the ``clustering signal'' and ``shot-noise'' components of the power spectra, which are relevant at low and high spatial frequency $k$ (Mpc$^{-1}$) respectively \citep{Bernal_2022}. In this way, the shot-noise is dominated by the contributions of the brightest galaxies (hence differing luminosity models giving a range of power spectra predictions), whilst the clustering signal is where galaxies below the detection limit can be probed \citep{Marcuzzo_2025}. 
Aside from comparing the spectra of our cubes to previous works, we can also compare it to sensitivity limits from \cite{CCAT_Prime_Collaboration_2022}. Due to the intensity cube units of Jy\,sr$^{-1}$, $P(k)$ has units Mpc$^3$\,(Jy\,sr$^{-1}$)$^2$. By convention, we present the power spectrum as $k^3P(k)/2\pi^2$ with units (Jy\,sr$^{-1}$)$^2$, which we plot against spatial frequency $k$.

We calculated the power spectra as in C24: Fast Fourier Transforming (FFT) the intensity cube, then taking the volume-normalised spherical average around the central location in Fourier space. We used shell width $\Delta k=0.3$\,Mpc$^{-1}$ as a trade-off between being able to separately trace the different regimes of power spectra (shot-noise and clustering signal dominated) and maximising signal-to-noise ($S/N$). In our calculations, the maximum $k$ value corresponded to the scale of one beam FWHM in order to avoid the range affected by beam attenuation. 

Regardless of the type of cube we take the power spectra of, we always assumed that the comoving volume of the intensity cubes corresponds to $z_\textrm{[CII]}$ for that frequency band \citep{Bernal_2022, Sato_Polito_2023}. This normalisation was to ensure consistency with power spectra and sensitivity limits from the literature. 
\subsection{Masking}\label{sec:methodMasks}

In order to clean CO emission from the intensity cubes, we tested the appropriateness of two masking techniques (``targeted'' and ``blind'') for our intensity cubes, and by extension the E-COSMOS field. These masks block out sources of CO signal whilst retaining the [CII], so that the total power spectra is dominated by [CII] signal. We wanted to avoid removing voxels with high [CII] intensity which may overlap with voxels with high CO intensity (``overmasking''), suppressing the [CII] power spectra. Testing this is vital, because if this partial or full overlap between CO or [CII] is found in the FARMER LP cubes, it may be reflected in actual EoR-Spec observations of the same field. 
When calculating the power spectra for a masked cube, we altered the FFT volume normalisation by dividing by an additional factor:
\begin{equation}
\textrm{Factor}=\left(\frac{V_\textrm{total}-V_\textrm{mask}}{V_\textrm{total}}\right)^2,
    \label{eq:MaskVolNorm}
\end{equation}

\noindent where $V_\textrm{total}$ is the total comoving volume of the intensity cube, and $V_\textrm{mask}$ is the volume which has been masked out. This ensures the FFT of the masked cube is appropriately normalised, corresponding to the suppression of Fourier amplitudes by masking.
\subsubsection{Creating and applying targeted masks}\label{sec:methodMasksTarget}

The targeted masking technique (\citealt{Viero_2019}, \citealt{VanCuyck_2023}, \citealt{Karoumpis_2024}) assumes that the majority of contaminant emission is emitted by known bright galaxies at lower redshift than the target. Consequently, galaxy catalogues would already include these galaxies, so we can use their data to remove contaminant line emission. This technique is convenient for EoR-Spec as the Euclid deep field catalogues cover E-CDF-S and COSMOS2020/5 cover E-COSMOS. 

\begin{figure*}[t]
\centering
\begin{tikzpicture}[node distance=1.75cm]

\node (start) [startstop] {Start};
\node (takeall) [io,right of=start, xshift=1.3cm] {Take combined CO intensity cube (set as ``base cube''), the [CII] and total cubes, and the CO masks.};
\node (checkmasksapplied) [decision, right of=takeall,xshift=2.95cm] {Have all masks been applied, determining the optimal masking order?};
\node (checklinesapplied) [decision, right of=checkmasksapplied,xshift=2.85cm] {Have all masks been tested on the base cube?};

\node (applymask) [process,right of=checklinesapplied, xshift=2.2cm] {Apply the mask to the base cube.};
\node (eparam) [process,below of=applymask, yshift=-1.9cm] {Calculate the efficiency parameter of the applied mask.};
\node (newbase) [process,left of=eparam, xshift=-2.2cm] {Apply mask with the highest efficiency parameter, remove it from pool of masks};
\node (savemap) [process,left of=newbase, xshift=-2.85cm] {Save the masked cube as the new base cube, calculate power spectrum.};
\node (optorder) [io,left of=savemap, xshift=-2.75cm] {Apply optimal masking order to [CII] and total intensity cubes, find cubes and power spectra};
\node (stop) [startstop,left of=optorder, xshift=-1.3cm] {End};

\draw [arrow] (start) -- (takeall);
\draw [arrow] (takeall) -- (checkmasksapplied);
\draw [arrow] (checkmasksapplied) -- node[anchor=south] {no} (checklinesapplied); 
\draw [arrow] (checklinesapplied) -- node[anchor=south] {no} (applymask); 
\draw [arrow] (applymask) -- (eparam);

\draw [arrow] (eparam) -- (checklinesapplied);
\draw [arrow] (checklinesapplied) -- node[anchor=east] {yes} (newbase);

\draw [arrow] (newbase) -- (savemap);
\draw [arrow] (savemap) -- (checkmasksapplied);
\draw [arrow] (checkmasksapplied) -- node[anchor=east] {yes} (optorder);
\draw [arrow] (optorder) -- (stop);

\end{tikzpicture}
\caption{Flowchart showing how we apply targeted masking. We do this separately for each mask radius.}
\label{fig:FigFlowchart_TargetedMasking}
\end{figure*}
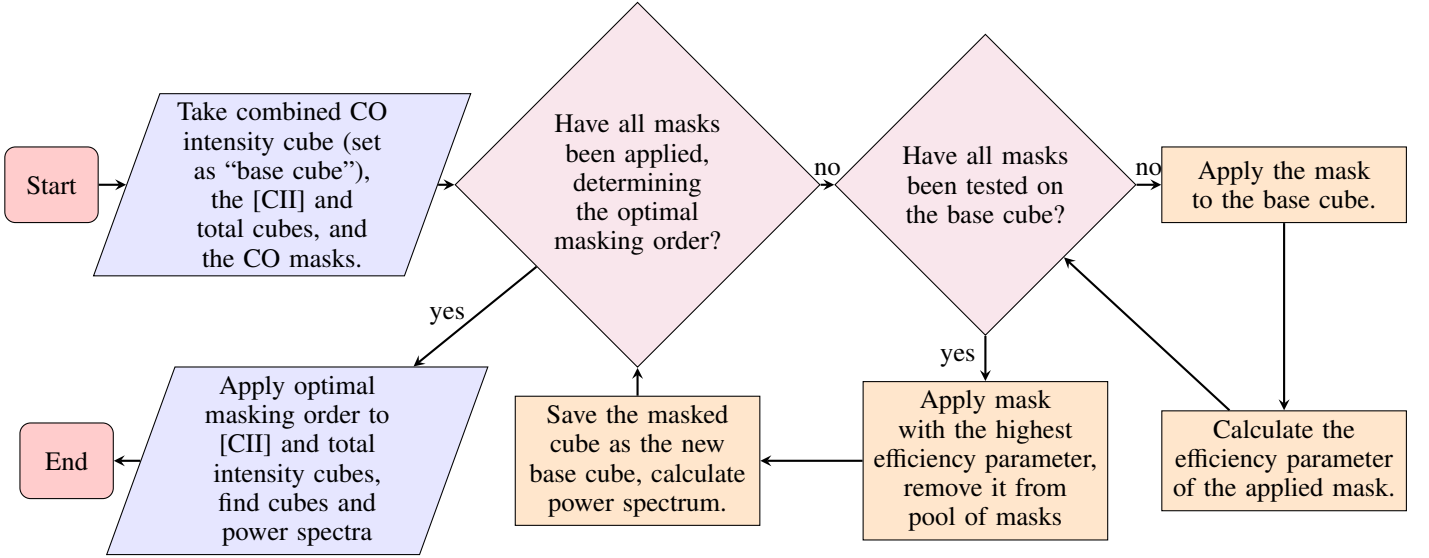

Our targeted masking procedure is summarised in Fig. \ref{fig:FigFlowchart_TargetedMasking}. We take the combined CO cubes and iteratively apply all masks to them. For each step, we find the optimal mask via an efficiency parameter, and repeat until all masks are applied, forming the optimal masking order which we subsequently apply to our cubes.

In our implementation we removed voxels one CO transition at a time, so we can identify which transitions contribute the most to the total CO line emission. To make these masks for each frequency range, we included bright galaxies corresponding to each transition in a mock external catalogue, whose locations we intended to mask. By default we only included non-extrapolated galaxies above the COSMOS2020 stellar mass limits in Table \ref{table:CANDELSlims} (similar to \citealt{Bethermin_2022}). We also used lower limits for galaxy selection below $z<0.5$, 
determined empirically to optimise CO removal whilst minimising masked volume, scaling linearly down to $10^6 M_\odot$ for $z=0.1$
. We exclude all galaxies at $z<0.1$ from our cubes, a consequence of the $L_\textrm{IR}$-mass relationship at low $z$ (Appendix \ref{appendix: low z}). In this way, the external catalogue contains galaxies from COSMOS2020 with high stellar mass, which we expect to contribute more to the CO intensity.

\begin{figure}[t]
 \centering
 \includegraphics[width=\linewidth]{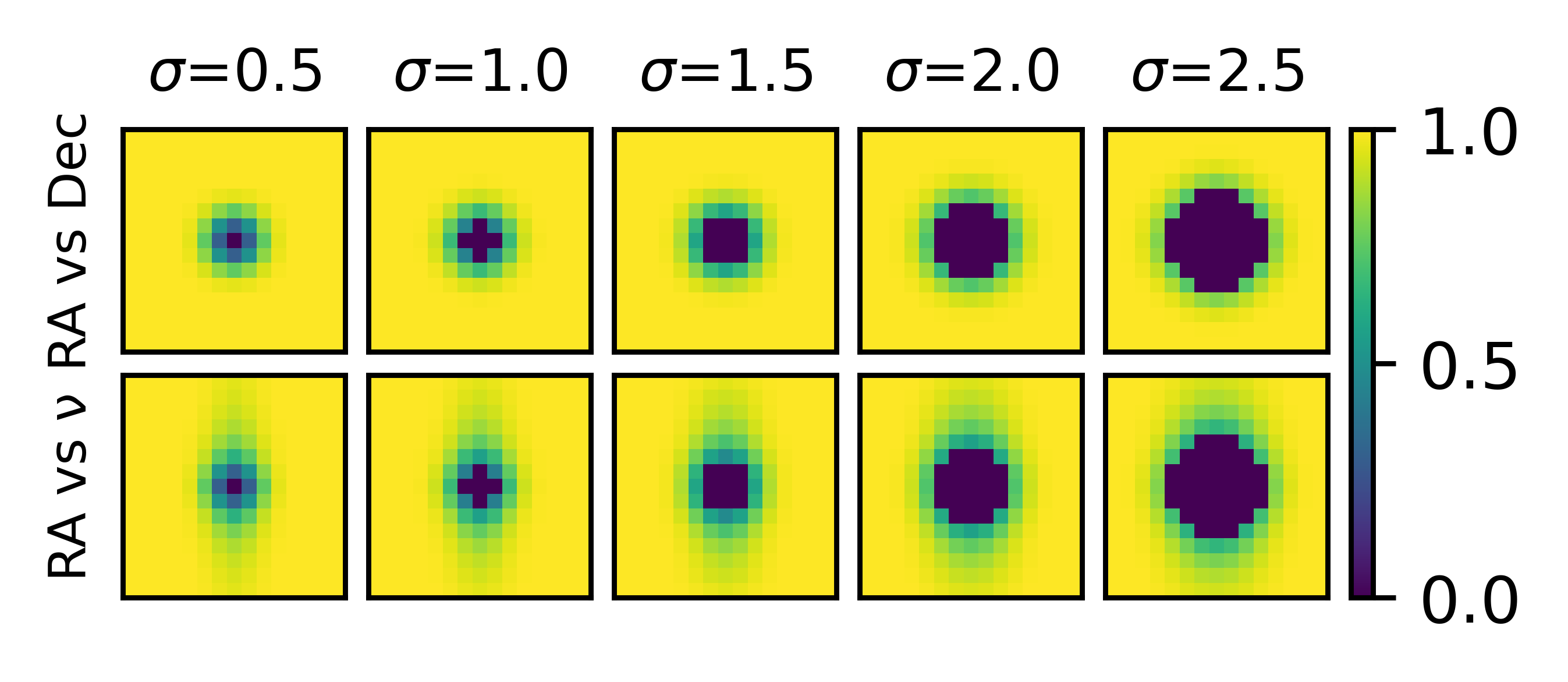}
 \captionof{figure}{Different radii masks, from $\sigma=0.5$-$2.5$ subdivided voxels. Upper subplots show the profile on the map (Gaussian), lower subplots show the frequency space profile (Lorentzian). We convolve these with galaxy locations, so locations marked as 1 are unmasked, locations marked as 0 are fully masked.}
 \label{fig:MaskTypes}
\end{figure}

After defining the external galaxy catalogue, we created masks with sizes optimised to remove CO intensity with minimal impacts to [CII]. Mathematically, these masks are 3D arrays with the same dimensions of the intensity cubes, with values between 0 and 1. In this way, multiplying these arrays with the intensity cubes is equivalent to applying the masks: locations with a value of 1 are kept the same, locations with values of 0 are fully masked, and other locations are proportionally suppressed. In constructing the mask shapes, we combined the methodologies of \cite{VanCuyck_2023} and \cite{Karoumpis_2024}. Following the former, all voxels within a certain radius of the galaxy locations are fully masked (set to 0) in a binary mask, this radius determined by the beam standard deviation ($\sigma$) in map- and frequency-space. While \cite{VanCuyck_2023} apodised this cube using POKER pipeline \citep{Ponthieu_2011}, we instead convolved the mask with the same 2D Gaussian beam/1D Lorentzian frequency profile used by observations following \cite{Karoumpis_2024}. This can be conceptualised as an inverted beam, fully masking the brightest points whilst partially masking the emission tail. 
We used five variations for the binary masks with different radii ($0.5\sigma$, $1\sigma$, $1.5\sigma$, $2\sigma$, and $2.5\sigma$), and in the masking procedure we kept each version separate. These convolved maps are referred to by their radii. The binary mask width determines the proportion of intensity removed from any given galaxy, from $\sim$20$-$80\%, indicating that wider masks are useful when more CO removal is required. However, wider area coverage introduces the risk of overmasking, as discussed in Section \ref{sec:results_targetmasking}. The mask shapes are shown in Fig. \ref{fig:MaskTypes}.

Treating each mask radius separately, we made two masks for each CO transition as in \cite{Karoumpis_2024}: one based on all galaxies from the external catalogue (``complete''), the other only including those above the galaxy main sequence (``bright''). The main sequence parameters were determined empirically with FARMER LP for each $\Delta z=0.5$ interval (Eq. \ref{eq:GalMainSequence}, Table \ref{table:CANDELSlims}). Then we found an optimal masking order, that is which masks remove the most CO per unit volume. We independently applied each bright and complete mask to the combined CO intensity cube, to find which maximise the efficiency parameter:
\begin{gather}
E=\frac{\textrm{STD}_{\textrm{before masking}}-\textrm{STD}_{\textrm{after masking}}}{\textrm{Number of newly masked voxels}},
\label{eq:Efficiency}
\end{gather}
\noindent where $\textrm{STD}$ is the standard deviation of all the values in the 3D arrays before and after masking. After finding the most efficient mask of that round, this process is repeated iteratively until all masks are applied. While we will not have the CO cube used to calibrate this efficiency order from our actual observations, as we can only observe the signal of the total summed intensity cube, we can make a first order approximation of this cube from our foreground catalogue. Typically, we found that bright masks were more efficient than complete masks as they cover the brightest sources in the fewest number of voxels, so our process usually applied them before their complete counterparts. We only included CO transitions up to $J$\,=\,9$-$8 in this procedure to mitigate overmasking, as our testing showed that masking higher transitions provides diminishing returns (see Section \ref{sec:results_targetmasking}). Once the optimal order is determined, we apply it to the total and [CII] cubes to find the impact of these masks on their signal. We show our results for these masks in Section \ref{sec:results_targetmasking}.

\subsubsection{Creating and applying blind masks}\label{sec:methodMasksBlind}

The blind masking technique, 
initially investigated in context of the frequency coverage of TIFUUN, 
aims to remove CO signal without relying on an external foreground galaxy catalogue, and so is useful for fields where this is incomplete. This is relevant for E-COSMOS in particular as it is only partially covered by COSMOS2020. Our technique also ignores the selection bias from a foreground catalogue and can include bright galaxies that it missed, represented in our simulations by theoretical extrapolated galaxies. It works because a galaxy at a given redshift will emit multiple lines with different rest (and so observed) frequencies, thereby correlating across frequency channels (also the driving principle behind cross-correlation spectra). For our given bands we expect to be able to observe various CO ladder transitions from the same galaxies, so if we can identify these bright sources by pairing them across map and frequency space, we can isolate and remove them. 

\begin{figure*}[t]
\centering
\begin{tikzpicture}[node distance=1.75cm]

\node (start) [startstop] {Start};
\node (taketomog) [io,right of=start,xshift=2cm] {Take total intensity cubes from relevant bands};
\node (localmax) [process,right of=taketomog,xshift=2.5cm] {Identify ``local maxima'' brightest voxels for all cubes for given \% cutoff};
\node (itcheckvoxels) [process,right of=localmax,xshift=2cm] {Compare and pair all maxima voxels based on map location};

\node (checkedallvoxels) [decision, below of=itcheckvoxels,yshift=-1.4cm] {Have all pairs of voxels been checked?};

\node (matchfreq) [process,left of=checkedallvoxels,xshift=-1cm, yshift=-2.2cm] {Iteratively apply CO/[CII] line freqs. to the pair, calculate redshifts of these maxima};

\node (redmatch) [decision,left of=matchfreq,xshift=-2.75cm] {Do the redshifts match for any set of lines?};

\node (ifCII) [decision,below of=matchfreq,yshift=-2.75cm] {Is one of the lines that match [CII]?};
\node (CIIlumcheck1) [process,left of=ifCII,xshift=-2.75cm] {Calculate luminosity of maxima based on flux and redshift};
\node (CIIlumcheck2) [decision,left of=CIIlumcheck1,xshift=-2.75cm] {From luminosity of [CII] and other line, is the [CII] assignment realistic?};

\node (notinclude) [process,above of=CIIlumcheck2, yshift=2.75cm] {Do not include pair in masking};
\node (include) [process,below of=CIIlumcheck1, yshift=-0.53cm] {Include pair in masking};

\node (mask) [process,below of=checkedallvoxels,yshift=-2.25cm,xshift=2cm] {Make masks out of all included pairs};

\node (end) [startstop, below of=mask,yshift=-1.5cm] {End};

\draw [arrow] (start) -- (taketomog);
\draw [arrow] (taketomog) -- (localmax);
\draw [arrow] (localmax) -- (itcheckvoxels);
\draw [arrow] (itcheckvoxels) -- (checkedallvoxels);
\draw [arrow] (checkedallvoxels) --  node[anchor=east] {no} (matchfreq);
\draw [arrow] (matchfreq) -- (redmatch);
\draw [arrow] (redmatch) -- node[anchor=east] {yes} (ifCII);
\draw [arrow] (redmatch) -- node[anchor=north] {no} (notinclude);
\draw [arrow] (ifCII) -- node[anchor=east] {no} (include);
\draw [arrow] (ifCII) -- node[anchor=north] {yes} (CIIlumcheck1);
\draw [arrow] (CIIlumcheck1) -- (CIIlumcheck2);
\draw [arrow] (CIIlumcheck2) -- node[anchor=east] {yes} (notinclude);
\draw [arrow] (CIIlumcheck2) -- node[anchor=east] {no} (include);
\draw [arrow] (include) -| (checkedallvoxels);
\draw [arrow] (notinclude) |- (checkedallvoxels);
\draw [arrow] (checkedallvoxels) --node[anchor=west] {yes} (mask);
\draw [arrow] (mask) -- (end);

\end{tikzpicture}
\caption{Flowchart of the steps for blind masking, for all the total summed intensity cubes we made. We repeat this procedure for every \% cutoff.}
\label{fig:FigFlowchart_Blindmasking}
\end{figure*}
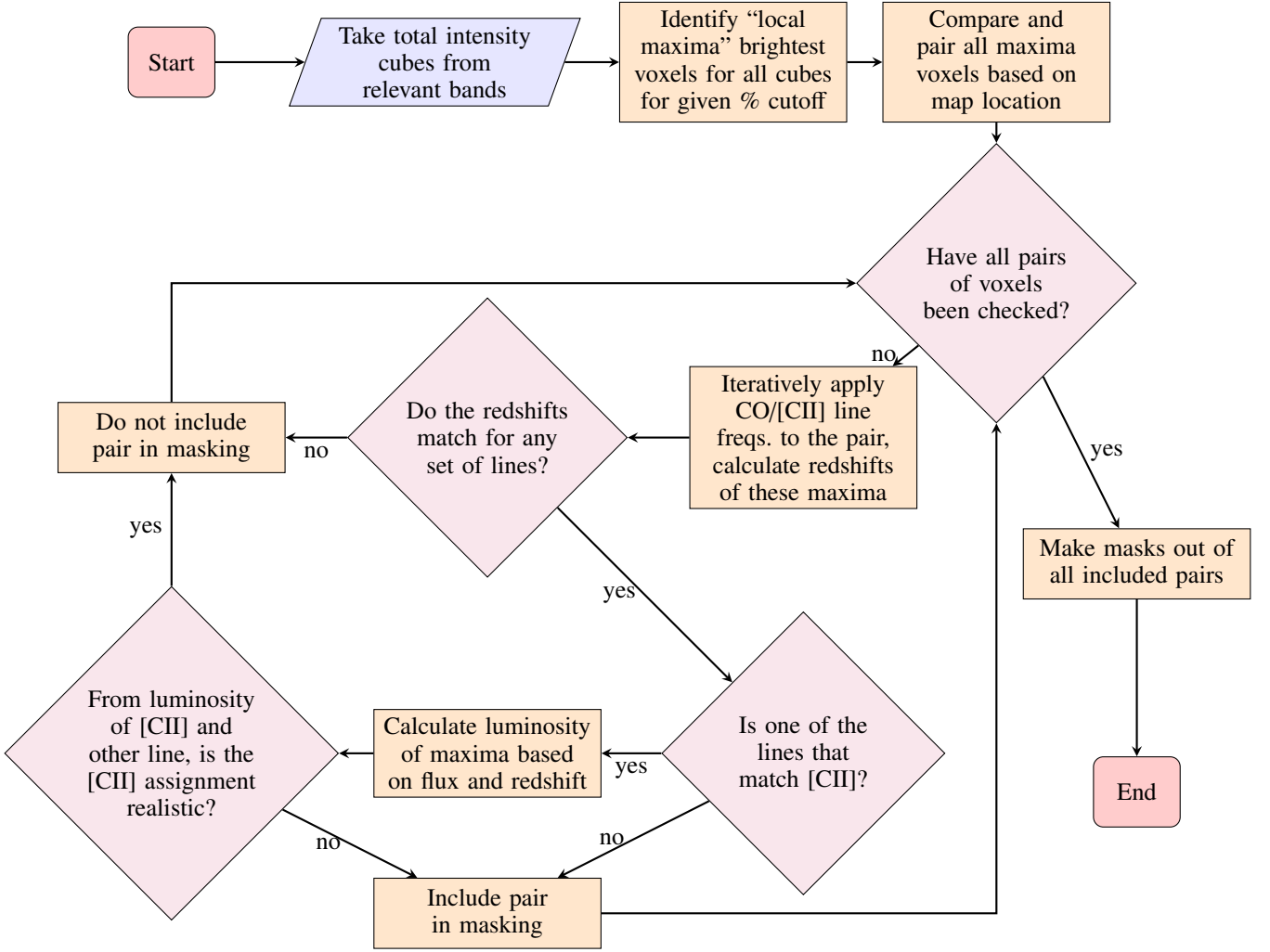

Our blind masking procedure is summarised in Fig. \ref{fig:FigFlowchart_Blindmasking}. We take the total summed intensity cubes for each frequency range, identify the local maxima within a given intensity cutoff, and paired them based on map location within the radius of one beam FWHM. For each voxel pair, we check if their frequencies could correspond to a single galaxy emitting in two different lines. If so, we then designate them as locations to be masked. However, we make sure to verify that the bright voxels do not correspond to [CII], or that any voxel identified as [CII] is not a false positive. We mask in order of the intensity cutoff.

We first identified the $x\%$ brightest voxels of the total intensity cubes across all intensity cubes, with $x$ varying from $0.1$ to $10\%$, as an intensity cutoff. As the brightest voxels of a cube are often adjacent due to beam spread, we only included the local maxima. For each percentile range, we compare voxels across all cubes and pair those within one resolution element
. This indicates the same galaxy emitting across multiple lines with different observed frequencies. Looking at both voxels in a pair, we used the observed frequencies to calculate redshifts for [CII] and the CO transitions. If the redshift intervals of the given voxels overlap by more than 50\% for any pair of spectral lines, they are paired and assigned the appropriate emission lines. If a voxel was identified as being dominated by CO emission it was masked, but if one of the voxels in the pair was [CII], we performed an additional check to make sure it was actually [CII] (and so to not accidentally keep CO line emission). We took the voxel intensities of those pairs and, taking those as the emission peak, calculated the $L_{\textrm{Line}}$ of the corresponding lines by inverting Eq. \ref{eq:IntensityGal} (using the median line emission models). Once these were determined, we checked if the estimated luminosities lay within 2\,dex of the FARMER LP line of best fits between $L_\textrm{[CII]}$ and $L_\textrm{CO (other)}$, to make sure the voxel luminosities are not physically impossible. These best fit parameters were determined empirically for FARMER LP over all redshift intervals
. If this condition was met, the [CII] was declared valid and so the voxel was not masked, else we masked it anyway. These fits served well as a quick automated check, and for actual observations each pair could be manually verified. We do this procedure with and without the lower frequency bands, as they would belong to a successor to EoR-Spec. The latter case would have fewer pairs, and so would be less efficient at masking.

After determining the locations, we made masks as in Section \ref{sec:methodMasksTarget} and applied them in order of masking depth $x\%$.
\subsection{White and correlated noise} \label{sec:methodNOISE}

To estimate the impact of noise over the proposed observing period, in some of our analysis we implemented a combined white and correlated noise model to find its impact on the power spectra. While past work such as \cite{Roy_2024} implemented Gaussian white noise derived from the projected $P_\textrm{N}$ described by \cite{CCAT_Prime_Collaboration_2022}, this assumes that atmospheric conditions at the FYST site follow \cite{Radford_2016} and \cite{Cortes_2020}, the detector assumptions are valid \citep{Choi_2020}, and there are no structural components to the noise. However, realistic observations will contain additional correlated noise components including inhomogeneous water vapour distribution, atmospheric turbulence which evolves over the sky and time, and scanning effects. The white noise power spectra does not include scale-dependent effects.
 
Time-ordered noise simulations can account for these components, and so have been developed for EoR-Spec using the Time Ordered Astrophysics Scalable Tools (TOAST) framework\footnote{\url{https://github.com/hpc4cmb/toast/tree/master} (\citealt{Errard_2015}, \citealt{Kisner_2023}).} (Dev et al. in prep). They expect an additional correlated noise factor taking the form of a scaling $1/f$ factor in our noise spectra, where $f$ is temporal frequency of the observing stream. These, alongside spatial components introduced from the scan patterns, can be mitigated by principle component analysis (PCA) cleaning techniques. In their simulations they made $4\textrm{\,deg}^2$noise cubes for the 350\,GHz band for a fraction of the proposed observing time (25\% of data with 50\% efficiency, so 1/8\,th total sensitivity). For this, they applied PCA cleaning by removing the 3 leading principal components which contain the dominant correlated modes. They also provided a version with purely white noise. We took a $1.44\textrm{\,deg}^2$ cutout of each map, dividing them by $\sqrt{8}$ and $\sqrt{2^2/1.2^2}$ to account for full observing time and the reduced map size. Noise spectra from these have shot-noise on the same order of magnitude to the Gaussian noise estimates, but due to the scale-dependency of the correlated noise, they have greater signal at low $k$ modes associated with clustering ($k<0.5\,\textrm{Mpc}^{-1}$). These simulations are not directly applicable to other frequency bands, due to different atmospheric opacity, but to make first order estimates we took different cutouts and rescaled the spectra using the predicted $P_\textrm{N}$ (Table \ref{table:FreqData}). 

When just analysing the impact of noise on the power spectra, we use the analytical formulation for sensitivity as below:
\begin{equation}
    \sigma(k)= \frac{P(k)+P_\textrm{N}(k)}{\sqrt{N_{\textrm{modes}}(k,\Delta k)}} \frac{1}{W(k)},
\label{eq:Sensitivity}
\end{equation}

\noindent where $\sigma(k)$ is the effective sensitivity, $P(k)$ is the total signal spectrum, and $P_\textrm{N}(k)$ is the noise spectrum (\citealt{Li_2016}, \citealt{Chung_2020}, \citealt{Karoumpis_2021}), which can modulated by the window normalisation factor $W(k)$ (defined for Prime-Cam by \citealt{Marcuzzo_2025}) to give signal-to-noise:

\begin{equation}
    S/N(k)= \frac{P(k)W(k)}{\sigma(k)},
\label{eq:StN}
\end{equation}

which is relevant for power spectra recovery and targeted masking. However, to find the impact of noise on blind masking, it is more convenient to add the noise to the cubes directly and calculate the resulting noise+total signal power spectra, as there we see the impact of noise creating intensity spikes. 
\section{Results} \label{sec:results}

After constructing the combined intensity cubes, we analysed their average intensity and power spectra. We first examine estimates of CO signal, before comparing it to [CII] and noise, thus determining the feasibility of the different masking techniques. 
\subsection{CO intensity cubes, comparisons to [CII] and noise}\label{sec:results_CO}

\begin{figure}[t]
 \centering
 \includegraphics[width=\linewidth]{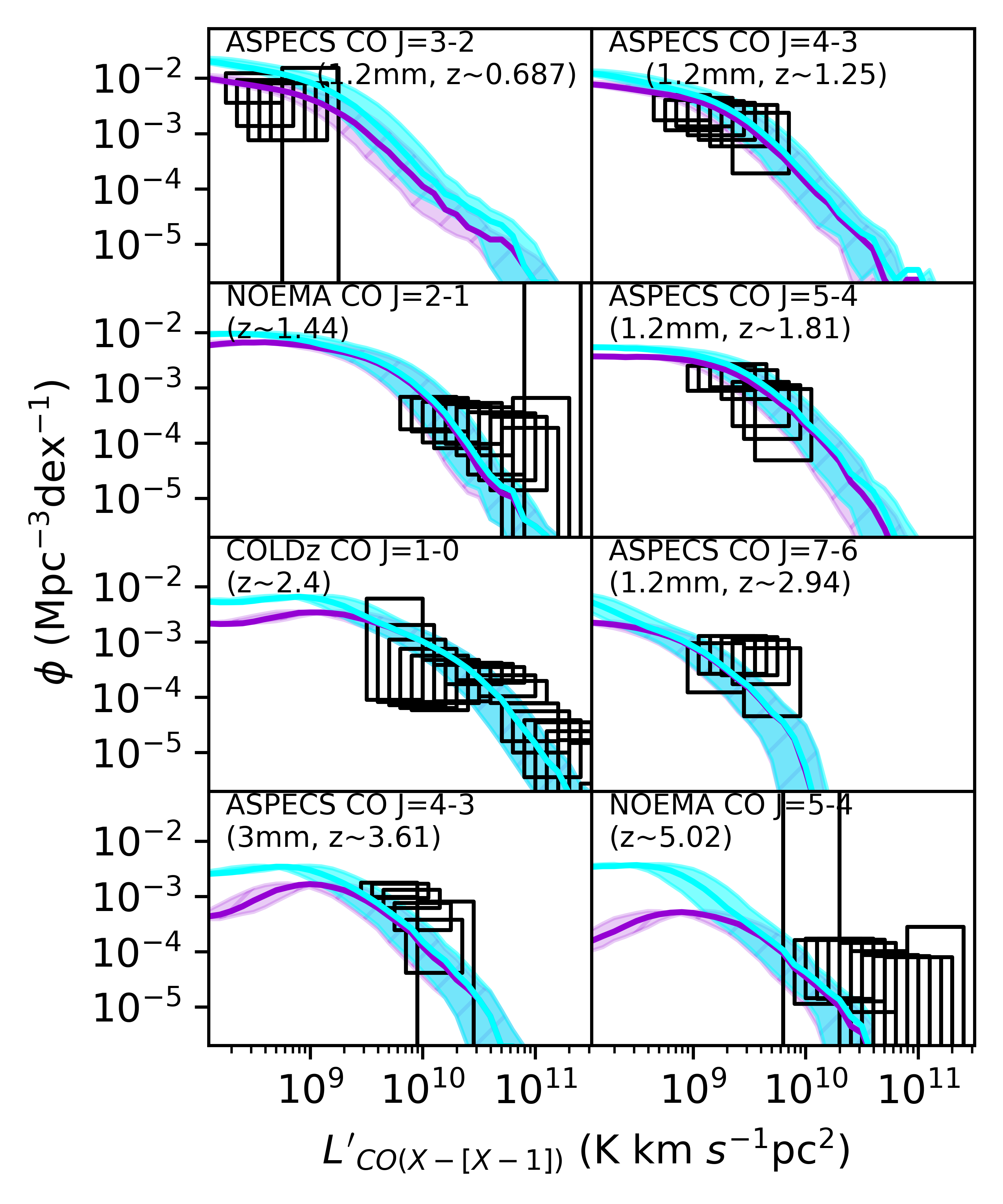}
 \captionof{figure}{Comparing CO line luminosity functions of FARMER LP to some from ASPECS/COLDz/NOEMA (\citealt{Decarli_2019,Decarli_2020}; \citealt{Pavesi_2019}, \citealt{Riechers_2019}; \citealt{Boogaard_2023}). $1\sigma$ errors for the observed samples are shown with black boxes, with shaded regions indicating FARMER LP (blue with extrapolation, purple/hatched without extrapolation, solid line indicating the Li model). Redshift of the relevant line for the survey is specified in each subplot. For direct comparisons we use $L'$ instead of $L$. A version including all luminosity functions is in Appendix \ref{appendix: Extra Figures}.}
 \label{fig:ASPECSComparison}
\end{figure}

We checked that our CO models were consistent with past observations by applying them to our samples, and then comparing the resulting luminosity functions to the ALMA study ASPECS \citep{Decarli_2019,Decarli_2020}, VLA study COLDz (\citealt{Pavesi_2019}, \citealt{Riechers_2019}) and observations using NOEMA \citep{Boogaard_2023}, the latter of which we refer to as NOEMA for simplicity. These deep-field surveys covered the Hubble Ultra Deep field for 100 and 250\,GHz, COSMOS and Great Observatories Origins Deep Survey-North (GOODS-N) from 30 to 39\,GHz, and Hubble Deep Field North for 100\,GHz respectively. For each CO line, we took FARMER LP galaxies corresponding to the ASPECS/COLDz/NOEMA observed frequencies (Eq. \ref{eq:FreqZRelation}), and applied our CO $J$\,=\,1$-$0 models and SLED template to this subsample to calculate the line luminosity functions (Fig. \ref{fig:ASPECSComparison}, Fig. \ref{fig:ASPECSComparisonFULL}). From these luminosity functions, FARMER LP lies within the $1\sigma$ uncertainty limits of ASPECS/COLDz/NOEMA for almost all transitions, with greater deviation when extrapolating at $z<1$ and $z>5$ due to the greater CANDELS factors (Section \ref{sec:methodSampleExtrap}). However, this analysis is a first-order estimate due to the impact of field-to-field variance, as ASPECS/COLDz/NOEMA are smaller fields which do not align with COSMOS2020 ($4.2\textrm{\,arcmin}^2$, $60\textrm{\,arcmin}^2$, $8.5\textrm{\,arcmin}^2$ and $1.44\textrm{\,deg}^2$ respectively, Section \ref{sec:discussion_masking}).

Despite these limitations we found that the line luminosity functions for the brighter CO models at $z<1$ reach the upper limits of ASPECS/COLDz/NOEMA, but at $z>3.5$ we find the opposite. The lower CO transitions at low $z$ contribute more total intensity than high transitions at higher $z$ due to the relative intensity contributions from our SLED (Appendix \ref{appendix: SLED}). For example, at 410\,GHz CO $J$\,=\,4$-$3 contributes 21\% to the total CO emission, whilst CO $J$\,=\,9$-$8 contributes 5\%. In this way we anticipate higher total CO emission than prior work, so our simulations should be viewed as challenging cases for [CII] recovery.

\begin{figure}[t]
 \centering
 \includegraphics[width=\linewidth]{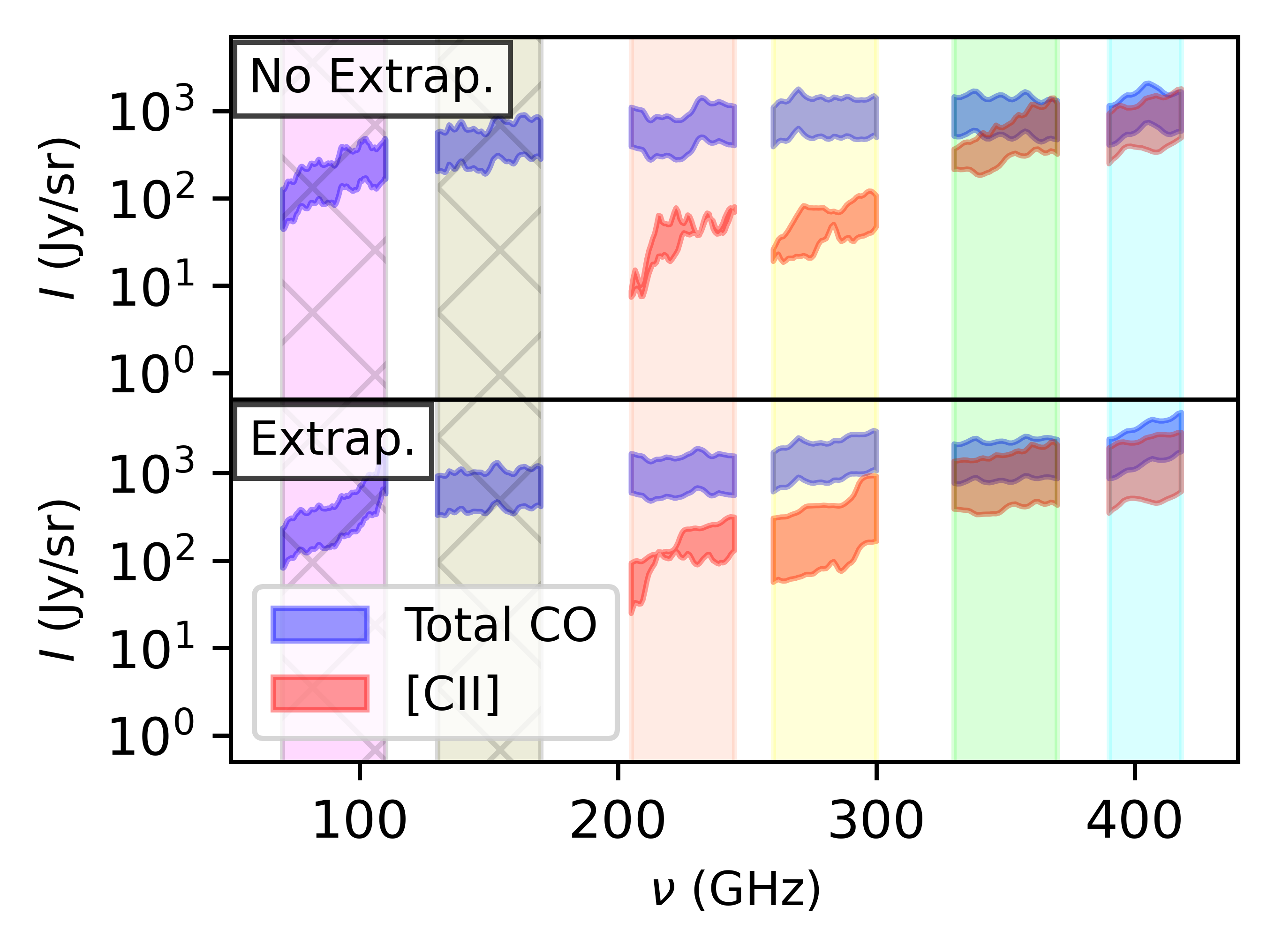}
 \captionof{figure}{Variation of the mean cube slice intensity with observed frequency (shaded bands being EoR-Spec and the extension), shown for [CII] (red) and the combined CO (blue), with or without extrapolation (upper/lower subplots). The average [CII] intensity decreases faster than CO as observed frequency decreases. The total CO and [CII] emission is equivalent for $>$300\,GHz, though for $<$300\,GHz CO dominates by more than $1/0.5$\,dex for the non-extrapolated/extrapolated case.}
 \label{fig:IvsFreq}
\end{figure}

To directly compare our predicted CO and [CII] emission, we juxtapose their average intensities for each frequency slice in our cubes (Fig. \ref{fig:IvsFreq}). They are equivalent above 300\,GHz with $\sim$$0.5\,$dex model variation, with a greater difference below 300\,GHz ($>$$0.5/1$\,dex with/without extrapolation). [CII] intensity follows the expected decrease for $<$300\,GHz, though the lower rate of decrease at 225\,GHz is likely due to the greater incompleteness and extrapolation variance at $z>6.3$ (C24). Even so, considering that CO intensity is greater by $>1$\,dex, cleaning at $225$\,GHz is unlikely to succeed with masking approaches considered here. This is also true for the bands below 200\,GHz, though as FARMER LP does not contain galaxy data at $z>10$ we cannot investigate this further with our method. Ultimately our results are consistent with \cite{Karoumpis_2024} and demonstrate the need for masking contaminants. 

\begin{figure*}[t]
 \centering
 \includegraphics[width=\linewidth]{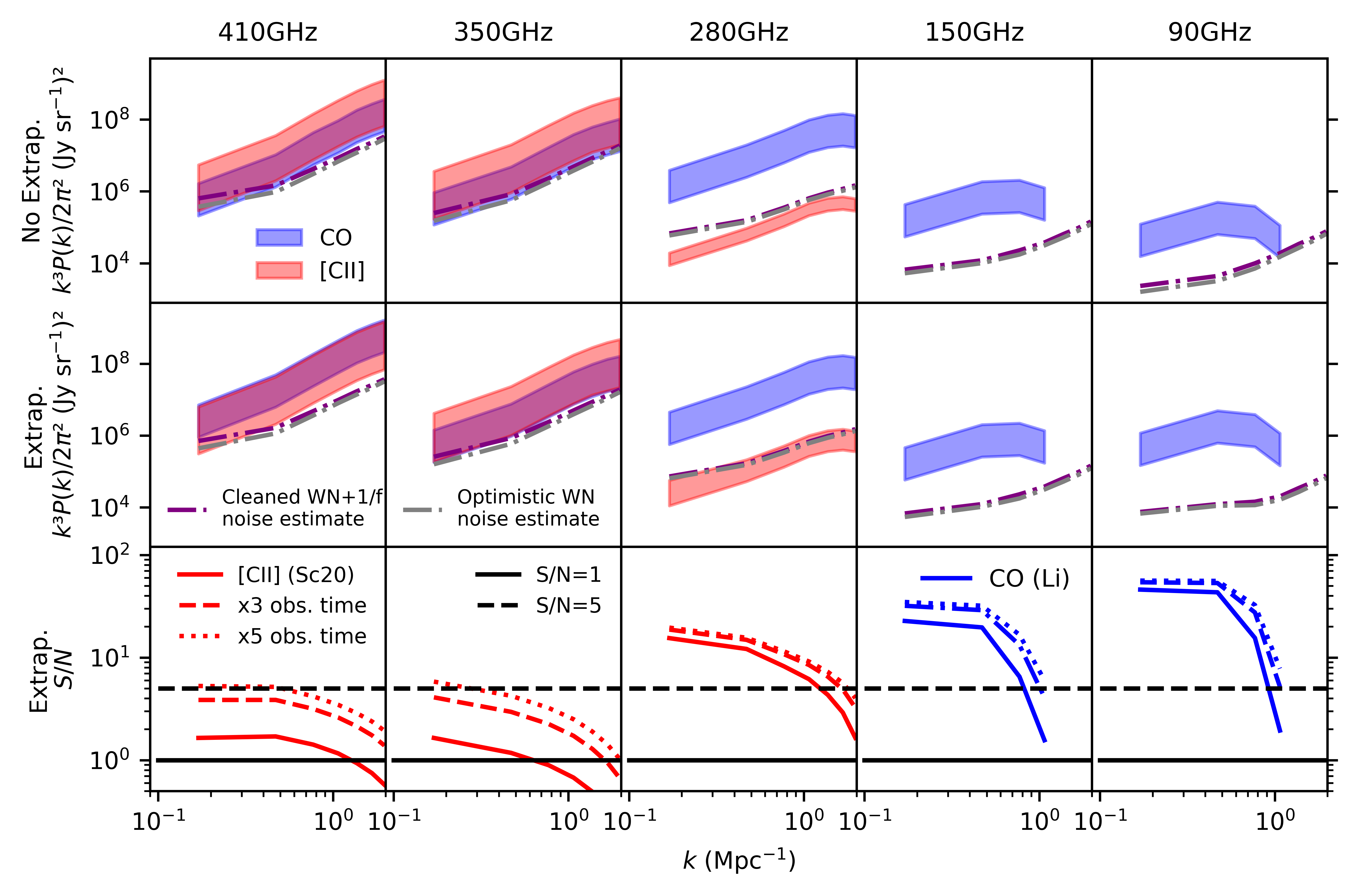}
 \captionof{figure}{Power spectra and $S/N$ of [CII] (red) and CO (blue) from the simulated cubes, for the 410, 350 and 280\,GHz EoR-Spec bands, and the 150 and 90\,GHz lower bands, using $k$ bins of width $\Delta k=0.3\,$Mpc$^{-1}$. Upper subplots do not include extrapolated galaxies, middle subplots include them. [CII] and combined CO power spectra have the same order of magnitude above 300\,GHz, with CO dominating below. CO is stronger in the extrapolated case, primarily due to increased CANDELS extrapolation at low $z$. Extrapolation has a larger impact on the 90\,GHz spectra ($1$\,dex) compared to 150\,GHz. Also included is the sensitivity $\sigma(k)$ determined using the Li and Sc20 models for median CO and [CII] signal (Eq. \ref{eq:Sensitivity}, dash-dot lines). We include the white plus correlated time-ordered simulation (purple) and a simpler white noise variation (grey), each normalised by full observing time, map size, and $P_\textrm{N}$. The cleaned WN+1/f sensitivities are used to calculate the $S/N$ ratios given in the lower subplots (Eq. \ref{eq:StN}), though the CO total is used for the lower bands. These use the median models for [CII] and CO, Sc20 and Li respectively. We also include examples when increasing observing times by 3 and 5, which are relevant for the higher frequencies to achieve $S/N=5$.}
 \label{fig:Pk_NoExtrap}
\end{figure*}

We also compare the relative strength of the CO and [CII] power spectra before masking (Fig. \ref{fig:Pk_NoExtrap}), including post-cleaning noise sensitivities for reference. As indicated by the average intensities in Fig. \ref{fig:IvsFreq}, the CO and [CII] power spectra are equivalent for 410 and 350\,GHz with 1\,dex model variation, but CO dominates at 280\,GHz. This shows the need for light and heavy masking respectively (Section \ref{sec:results_targetmasking}). Note that the brightest [CII] model (C24\_m3) exceeds the CO equivalent (SargMS) at 350\,GHz, despite having lower average intensity. This is because C24\_m3 produces a greater proportion of highly luminous galaxies, thereby increasing the intensity cube variance and so shot-noise.
All but the dimmest CO and [CII] models eclipse the predicted atmospheric sensitivity for all $k$ modes at 410 and 350\,GHz, though fainter models and those below 300\,GHz fail to do so for $k<0.5\textrm{\,Mpc}^{-1}$ (see Section \ref{sec:discussion_masking+noise}). This is even accounting for the shot-noise suppression introduced by the Lorentzian profile in frequency space (\citealt{Marcuzzo_2025}, Appendix \ref{appendix: Lorentzian V Gaussian}), which reduces $S/N$. However, to achieve an $S/N>5$ for the high frequency bands we would need additional observing time, on the order of five times to be safe. Regardless, it is reassuring that the extrapolated cases  
give $S/N$ which is concordant with the analytical treatment of C24. As this is peaks for low $k$ modes, it will be more feasible to recover the clustering signal.

From both statistics we found that CO emission drops more slowly than [CII] as frequency decreases, due to how each transition contributes to the total contaminant signal. Transitions from CO $J$\,=\,4$-$3 through $J$\,=\,7$-$6 are brighter than higher transitions following our SLED (Appendix \ref{appendix: SLED}), especially as those galaxies lie at lower $z$ with less distance to us and so produce greater intensities. As observed frequency decreases, these transitions contribute less intensity as they pass the 
CO luminosity density maximum at $1.5<z<3$ (\citealt{Decarli_2019,Decarli_2020}, \citealt{Pavesi_2019}, \citealt{Riechers_2019}), closely tied to the SFR maximum at similar redshift (``cosmic noon'', \citealt{Madau_2014})
. However, even as they leave the observed frequency range, new transitions (e.g. CO $J$\,=\,3$-$2) enter and contribute to the total observed CO. Despite this, we see a decrease in total CO emission as these new transitions have lower luminosity in our SLED. By contrast, [CII] does not have a multi-component nature so the emission simply falls as luminosity distance increases, following a $1/r^2$ power law. For 150 and 90\,GHz the CO signal eventually drops, as the brightest transitions pass beyond $z\approx2$ and we run out of transitions after CO $J$\,=\,1$-$0, which is up to 2\,dex fainter than CO $J$\,=\,7$-$6. This is reflected in the lower magnitudes of CO emission in Fig. \ref{fig:IvsFreq} and Fig. \ref{fig:Pk_NoExtrap}. Despite this, projected white plus correlated noise falls faster than total intensity for the majority of $k$ modes, so constraining CO will be feasible for observations made below $<300$\,GHz. In addition, we find that CANDELS extrapolation has a greater impact on 90\,GHz compared to other bands, as the observed band includes the brightest CO transitions at $z>5$ (with a CANDELS ratio of 0.5).

\subsection{Targeted masking}\label{sec:results_targetmasking}

From seeing the relative strength of CO and [CII] (Section \ref{sec:results_CO}), we only attempted to recover the [CII] signal by masking for $280$, $350$, and $410\,\textrm{GHz}$. We did not include $225\,\textrm{GHz}$ because [CII] intensity up to $1.5$\,dex weaker than CO, and is also biased by requiring extrapolating over a large area not covered by the UltraDeep slices (approximately 50\% of the field, C24). 

While we show the impact of targeted masking on CO, [CII], and combined power spectra, in actuality we will only observe the combined signal plus any noise we are unable to remove. The aim of masking is to infer [CII] from the total intensity cube, so we must get the masked [CII] and combined signals as close as possible. To show this, we plot the power $k^3P(k)/2\pi^2$\,$\textrm{(Jy/sr)}^2$ for a fixed $k$ mode ($k\approx0.15\,\textrm{Mpc}^{-1}$, $\Delta k=0.3\,$Mpc$^{-1}$) against the percentage of cube voxels masked. By default we only mask CO transitions up to CO $J$\,=\,9$-$8 to avoid overmasking (see Fig. \ref{fig:CompareKvVC2}).

\begin{figure*}[t]
 \centering
 \includegraphics[width=\linewidth]{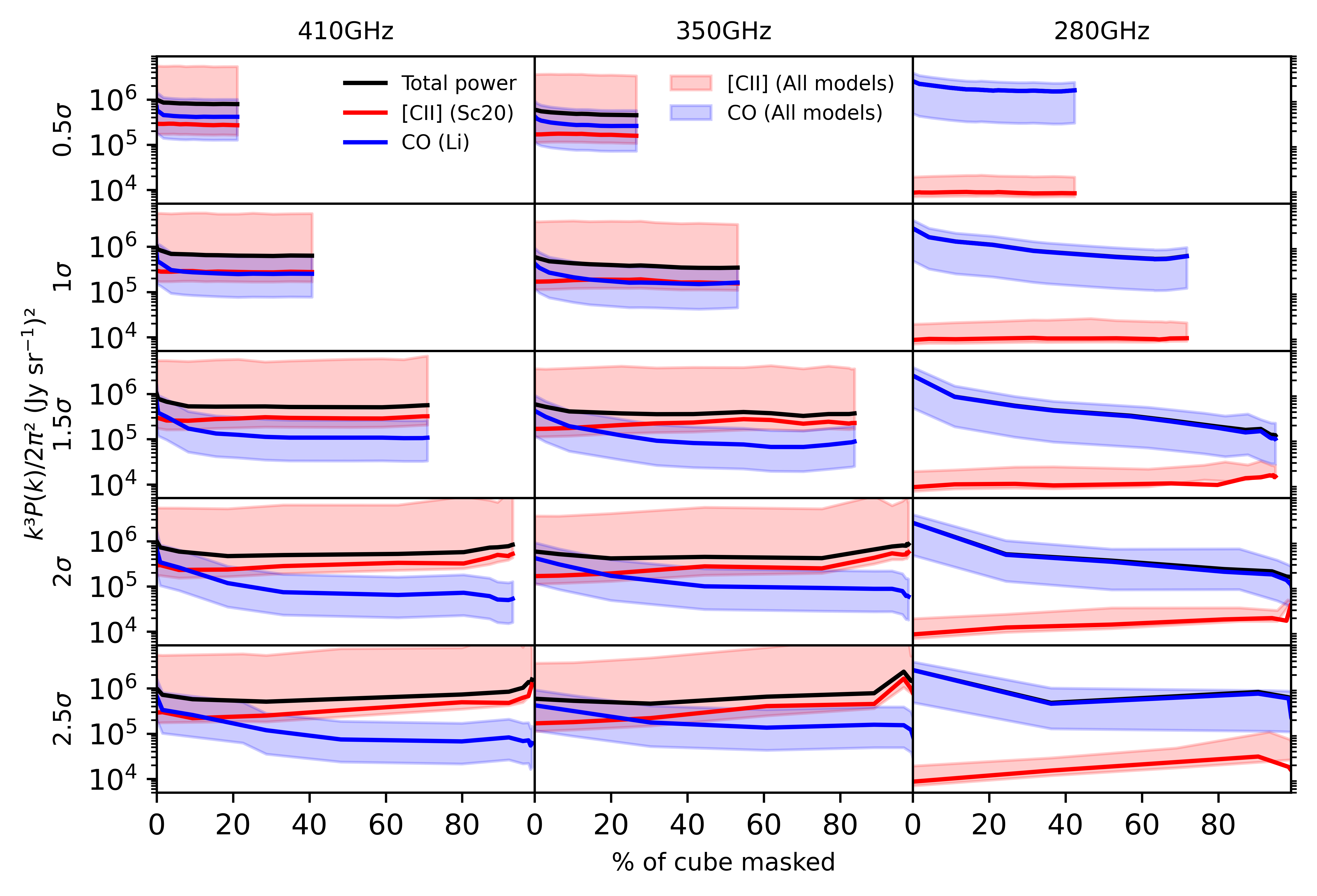}
 \captionof{figure}{Masking without extrapolation for cubes representing the 410, 350 and 280\,GHz bands, for all mask radii. CO is represented by blue, [CII] by red, combined by black, with solid lines representing the median models Li and Sc20 respectively. These illustrate how the power spectra at $k\approx0.15\textrm{\,Mpc}^{-1}$ ($\Delta k=0.3\,$Mpc$^{-1}$) changes as more cube is masked, with broader masks covering more volume but removing more CO. }
 \label{fig:CompareKvVC}
\end{figure*}

\begin{figure}[t]
 \centering
 \includegraphics[width=\linewidth]{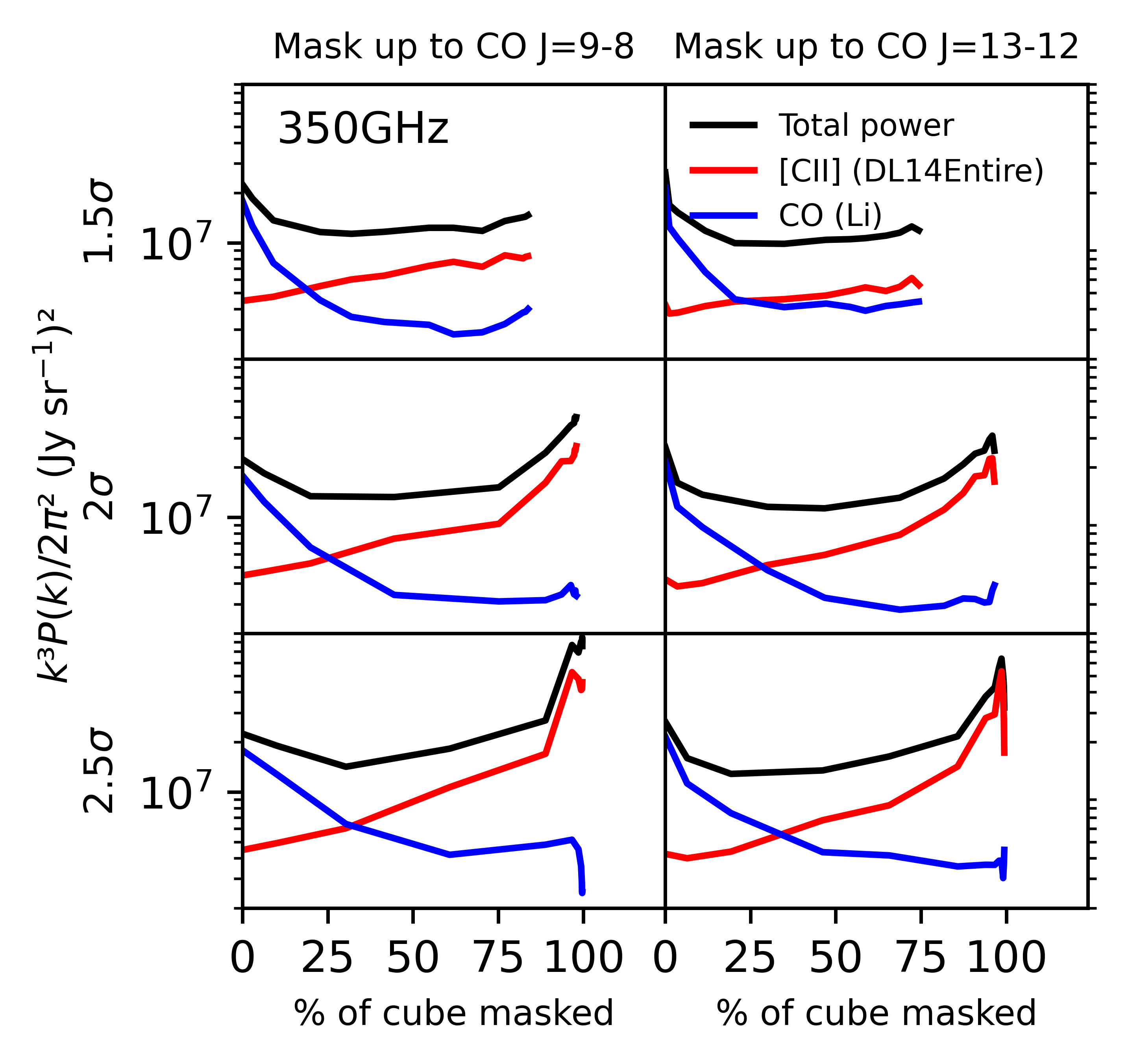}
 \captionof{figure}{
 Comparing masks ($1.5\sigma$, $2\sigma$, $2.5\sigma$; upper, centre, and lower subplots) when applied to a non-extrapolated cube with a low [CII] model (DL14 Entire) at 350\,GHz using the median CO Li, when masking CO transitions up to $J$\,=\,9$-$8 and $J$\,=\,13$-$12 (left and right subplots). This was for a $\Delta k=0.3\,\textrm{Mpc}^{-1}$ bin at $k=0.15\,$Mpc$^{-1}$. We find that masking up to $J$\,=\,13$-$12, combined with larger masks, approaches $100\%$ masking volume resulting in stronger target signal suppression. }
  \label{fig:CompareKvVC2}
\end{figure}

First we consider the scenario where the foreground catalogue used in targeted masking is complete (with or without extrapolation). Fig. \ref{fig:CompareKvVC} compares the cubes for the different radii masks at 280, 350 and 410\,GHz. As expected, the CO signal drops as the primary contaminant voxels are removed from the cubes, and the [CII] signal increases because of the reduction in the survey volume (decreasing the comoving volume normalisation factor, Section \ref{sec:methodMasks}). In addition, wider masks cover more comoving volume ($1.5\sigma$ covering more than thrice the volume of $0.5\sigma$). This is reflected in the signal cleaning, as the wider masks remove CO signal by up to 1\,dex due to their wider footprint, but $0.5\sigma$ masks do not clean more than $\sim$0.3\,dex. Consequently [CII] recovery using $1.5\sigma$ masks or wider is possible above 300\,GHz, with $<0.2$\,dex difference between the total and [CII] for the ideal model, and masks at $1\sigma$ or below may be feasible depending on the exact emission model. However, even $2.5\sigma$ masks fail to recover [CII] at 280\,GHz at maximum masking depth. This analysis assumes that noise is sufficiently removed for these cubes.

We also encountered overmasking, when the target [CII] signal is reduced due to accidental removal of [CII] sources in the process of cleaning CO, causing a tell-tale downward spike of total signal (up to $\sim$0.5\,dex). This is shown in Fig. \ref{fig:CompareKvVC2}, where we compare masking up to $J$\,=\,9$-$8 against $J$\,=\,13$-$12. We found that this only occurred for us when the masked volume approached 100\%, for which overmasking is inevitable, so masking the faintest CO transitions (above $J$\,=\,9$-$8) is typically counterproductive. It also means that we must be careful when selecting the width of masks. The exception to the above analysis is that we found overmasking when implementing the bright $J$\,=\,5$-$4 mask at 410\,GHz (Fig. \ref{fig:Blindwithnolowfreq}). Overmasking emerging in our COSMOS-based simulations indicates that this particular transition should not be masked in any actual observations of the E-COSMOS field. 
However, this being the only notable transition with overmasking is encouraging for targeted masking in E-COSMOS. 
In practice, iteratively testing different widths of masking (from wider to thinner), limited to the most important CO transitions, will be vital to optimise observations.

When the foreground catalogue does not include the extrapolated faint galaxy population present in the simulated intensity cubes, masking becomes less effective as some CO-emitting galaxies remain identified (Fig. \ref{fig:Blindwithnolowfreq}). In this scenario, the volume-normalisation correction can amplify the CO contribution in the recovered power spectra. While targeted masking does not always fail here, this shows that we need sufficiently complete foreground catalogues to be confident in our results, which is not true for the whole E-COSMOS field. Areas covered by more complete surveys such as COSMOS2025 (\citealt{Weibel_2024}, \citealt{Shuntov_2025}) avoid this issue, though their smaller $0.54\textrm{\,deg}^2$ sky coverage is more sensitive to cosmic variance (see Section \ref{sec:discussion_masking}). Consequently, we must find alternate solutions.
\subsection{Blind masking}\label{sec:results_blindmasking}

\begin{figure}[t]
 \centering
 \includegraphics[width=\linewidth]{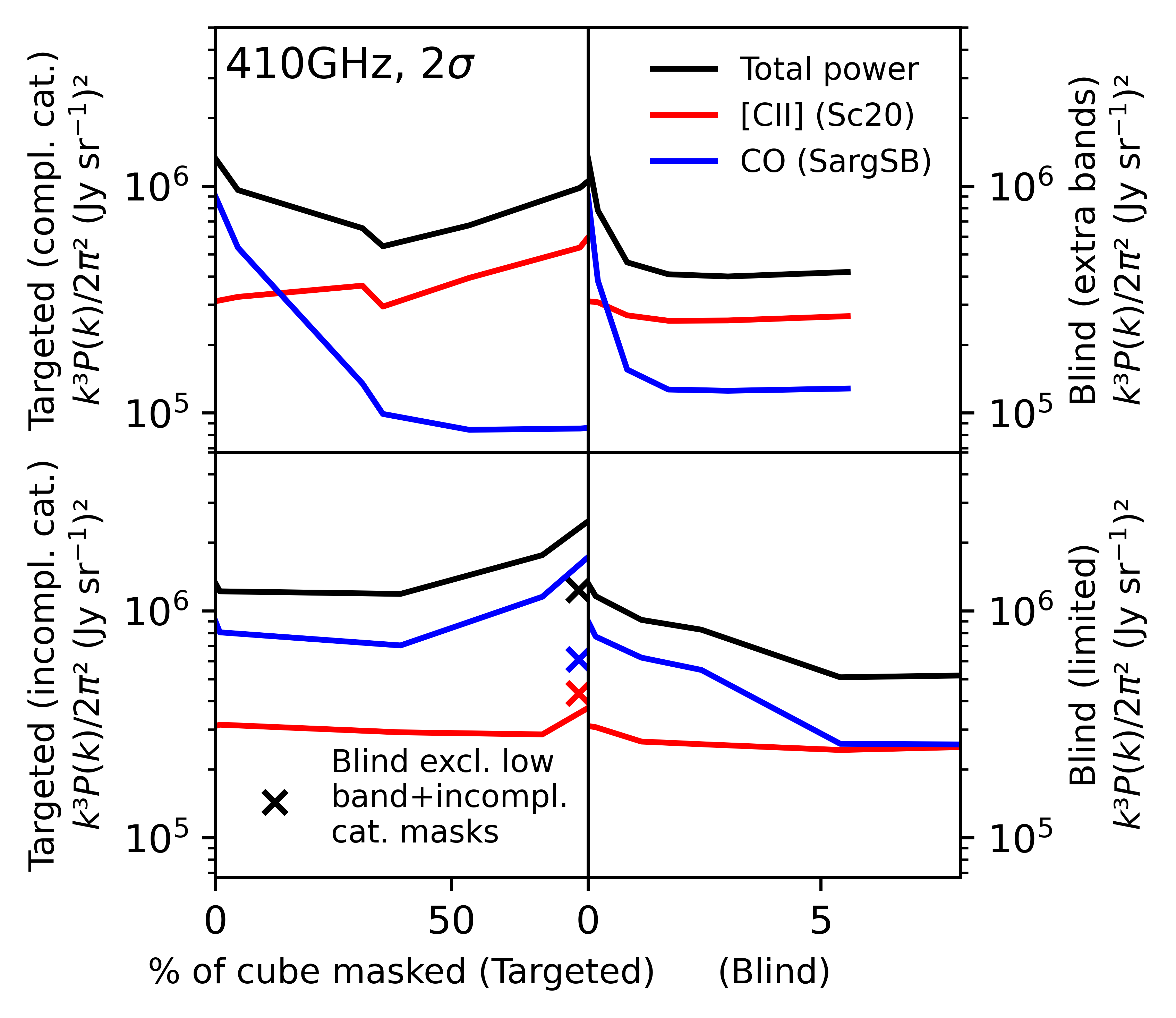}
 \captionof{figure}{
 Demonstrating targeted masking and blind masking on an extrapolated intensity cube with and without optimal conditions, where we would expect [CII] to dominate post-masking. This was for a $\Delta k=0.3\,\textrm{Mpc}^{-1}$ bin at $k=0.15\,$Mpc$^{-1}$. For all cases we use $2\sigma$ masks at 410\,GHz, Sc20 for [CII], SargSB for CO. Targeted masking is left, with top-left including extrapolated galaxies in the masking catalogue, and lower-left without a complete catalogue. Blind masking with/without the 90 and 150\,GHz bands is shown on the upper/lower-right. Both techniques succeed when including additional information, but are less successful without it by $1/0.2\,$dex respectively. For targeted masking, the CO power spectra rebounds due to volume correction. The crosses in the lower-right subplot show the results of combining the two incomplete techniques, giving an improvement of $0.5$\,dex.}
 \label{fig:Blindwithnolowfreq}
\end{figure}

We first investigated blind masking with extrapolated intensity cubes when including the hypothetical 90 and 150\,GHz bands in the procedure, to allow for additional voxel matching (Fig. \ref{fig:Blindwithnolowfreq}). This masks fewer voxels ($\approx10\%$ volume coverage in the 410\,GHz regime) when compared to successful targeted masking, so we see reduced volume correction effects. Despite masking less volume, the reduction of CO is the same order of magnitude as targeted masking, because blind masking exclusively targets the brightest voxels. Consequently, blind masking is superior if the catalogue is incomplete and noise is low (Section \ref{sec:discussion_masking+noise}). In addition, blind overmasking never exceeded 0.1\,dex, though it always occurred within the initial $3\%$ cutoff.

However, the technique is limited when only including the EoR-Spec bands. This is because some of the bright CO sources in those bands were only paired with voxels in the lower bands, resulting in them being missed and so a shot-noise discrepancy up to 0.5\,dex (Fig. \ref{fig:Blindwithnolowfreq}). Despite this, even when only using the EoR-Spec frequency range the technique is still more effective at recovering [CII] than targeted masking with an incomplete catalogue. Conversely, if other observing bands were included by cross-instrument analysis (such as with TIM or TIFUUN), we would expect blind masking to be improved in turn. We also note that blind masking and targeted masking complement each other and can be used in tandem, as while these masks trace the same underlying structure they have different biases. 

These results are all under the assumption that noise is reduced to the point of being negligible, or at least that we can successfully identify line emission peaks from contaminants and ignore intensity peaks that originate from noise. From applying cleaned and white correlated noise estimate to our intensity cubes, we found the technique is compromised by the higher noise levels above 300\,GHz. Aside from making the technique less effective for those specific bands (by up to 0.5\,dex, Fig. \ref{fig:RoyNoise}), those peaks being useless also reduces the effectiveness at lower bands as less pairing occurs. A full discussion of noise impacts is in Section \ref{sec:discussion_masking+noise}.

\section{Discussion} \label{sec:discussion}
\subsection{CO signal}\label{sec:discussion_CO}

Our technique successfully extended the methodology of C24 to include CO emission in intensity cubes using empirical data, with our predictions including conservative lower limits and extrapolated estimates for EoR-Spec which are consistent with ALMA, VLA, and NOEMA observations. As with [CII] models in C24, applying a wide range of models to the entire sample is better matched to the level of observational constraints than using more sophisticated galaxy models, to reduce the number of parameters and make the models appropriate for the data we have available. 
When extrapolating, the median and upper CO $J$\,=\,1$-$0 models provide greater CO power spectra shot-noise compared to the literature (\citealt{Bethermin_2022}, \citealt{VanCuyck_2023}, \citealt{Karoumpis_2024}), due to our model selection criteria and CANDELS extrapolation at $z<1$.  We see this even after removing outliers at $z<0.1$. Therefore, our models provide a pessimistic high CO-contamination scenario for recovering [CII] signal, and the successful masking is encouraging for the viability of our techniques. This is also why our lower limit (SargSB) aligns better with past work, compared to the median model (Li). 
We also implemented additional lower frequency ranges from \cite{Roy_2024} to aid blind masking and cross-correlation. As the bright CO 
line emission comes from higher redshift when observing with these additional bands, and our CO line luminosity functions are lower at high $z$, the forecasted power spectra act as lower limits for these bands. These bands allow for correlations with dedicated CO experiments at lower frequencies, such as COMAP \citep{Ihle_2019}, and additionally aid contaminant masking strategies.
\subsection{Effects and challenges of masking}\label{sec:discussion_masking}

The premise underpinning masking is that a significant fraction of the CO emission originates from a small number of bright galaxies. Our empirical modelling supports this picture, which can be demonstrated by considering the opposite scenario. If the CO signal were dominated by large populations of faint galaxies, masking a limited number of bright sources would have little effect on the recovered power spectra. However, the line luminosity functions derived from FARMER LP follow Schechter-like distributions with shallow faint ends below the knee (Fig. \ref{fig:ASPECSComparison}), indicating that the total CO emission and shot-noise are primarily driven by brighter galaxies. This supports the assumptions underlying masking for the FARMER LP-based simulations.

These masking principles are reinforced by the impact of extrapolation on the shot-noise. CANDELS extrapolation contributes $\sim$90\% to the shot-noise added by extrapolation, as it adds bright sources and so increases the second moment of the luminosity function while mass function extrapolation only adds faint galaxies. Even if we extended mass function extrapolation below the $<10^8 M_\odot$ limit we would not expect this to change, due to the aforementioned shallow gradient below the Schechter function knee, and in any case it would be unwise to extend it to an arbitrary depth at high $z$. Including mass function extrapolation is still important due to it bolstering the clustering signal, as these fainter sources are distributed around brighter ones (C24), and its contribution to the shot-noise is non-negligible. CANDELS extrapolation is strongest at $z<1$ and $z>3.5$, increasing the shot-noise of the lowest CO transitions and [CII]. Our results are supported by verifying the resulting mass functions against those of the COSMOS2025 catalogue, which had smaller sky coverage but deeper mass completeness (Appendix \ref{appendix: COSMOS2025Dev}). Consequently, our understanding of galaxy completeness at the high end of the mass function is crucial to determine limits on the CO power spectra, as discrepancies in the number of bright sources has the most impact on predicting shot-noise.

Our targeted masking was broadly consistent with prior work, being successful for 350 and 410\,GHz and less feasible for 280\,GHz, despite our high-CO models where the median CO power spectra typically exceeded [CII]. We found that this higher CO contamination made the subtler masking of $<1.5\sigma$ masks inappropriate for cleaning as they only removed $<40\%$ of the line emission of a given galaxy. Consequently, when treating real maps it will be safer to use broader masks first to ensure [CII] recovery, then implementing narrower masks if feasible.

If the foreground catalogue is incomplete then targeted masking becomes more challenging, as it fails to clean key bright CO sources (Section \ref{sec:results_targetmasking}). This is exacerbated for lower CO transitions at $z<1$, which are simultaneously brighter according to our SLED templates (Fig. \ref{fig:ASPECSComparison}, Appendix \ref{appendix: SLED}), but are also likely to be missed according to the low CANDELS ratios at that redshift (hence us extrapolating more bright galaxies there, Table \ref{table:CANDELSlims}). 
In the context of EoR-Spec, assuming our extrapolation method is accurate to first order as indicated by COSMOS2025 (Appendix \ref{appendix: COSMOS2025Dev}), the incomplete COSMOS2020 catalogue would be insufficient for recovering [CII] in E-COSMOS on its own. COSMOS2025 itself avoids this issue, but the smaller $0.54\textrm{\,deg}^2$ sky area coverage limits accessible $k$ modes and is subject to greater cosmic variance. To reliably recover [CII] power spectra via targeted masking it is necessary to achieve better coverage of E-COSMOS field at low $z$ from conventional surveys, or we would need to simultaneously use other cleaning techniques. As the Euclid E-CDF-S field will have equivalent depth to E-COSMOS (up to $26-27$\,mag, \citealt{Euclid_2025_main}), we anticipate similar conclusions for that region, although we cannot accurately test this with Quick Release 1 data. We also found that overcoming catalogue redshift uncertainty is needed to ensure effective masking, as these errors can result in accidentally masking the wrong voxels (Appendix \ref{appendix: RandomLocMasking}). Peak frequency matching or spectral line fitting could help mitigate these uncertainties. 

Fortunately, blind masking shores up the issues inherent to targeted masking by recovering signal when the foreground catalogue is incomplete, even when not including other frequency bands (from the hypothetical lower bands or other instruments, Fig. \ref{fig:Blindwithnolowfreq}). However additional bands will be required to definitively recover [CII], as otherwise $P_\textrm{[CII]}(k)/P_\textrm{CO}(k)\approx1$, limiting the applicability of blind masking in the early stages of EoR-Spec operation. This therefore justifies including lower bands in EoR-Spec below $<$200\,GHz, despite the reduced resolution of over 1\,arcmin$^2$. Correspondingly, cross-instrument observations that cover the same fields will also be able to enhance this technique, providing additional justification for instrument co-ordination beyond cross-correlation. However, in the short term, it will be necessary to combine targeted and blind masking for the best chances of [CII] recovery.  

For both techniques, we found overmasking was usually negligible unless masking approached 100\% volume coverage. To prevent this latter case, careful selection of the relevant CO lines and the mask widths by testing with galaxy catalogue data will be needed. Such testing can also verify any individual CO transitions that do overmask, in our case the CO $J$\,=\,5$-$4 bright mask for 410\,GHz. Due to cosmic variance statistics (see below) this should hold for the wider E-COSMOS field, but we cannot confirm this at present. This uncertainty will be less of an issue for E-CDF-S, as the Euclid deep fields will cover the entirety of it, so any modelling would be applicable to the entire E-CDF-S field.

Quantifying field-to-field variance is vital as the specific field determines the galaxy distribution, and so signal strength. COSMOS2020 covers one-third of the E-COSMOS field, so the structure we find has knock-on effects for E-COSMOS, as discussed by \cite{Keenan_2020} and \cite{Gkogkou_2023}. The latter found that the [CII] shot-noise of $1\textrm{\,deg}^2$ fields can vary up to 1\,dex as bright sources drift in and out of the map area (with $\sigma\approx0.25$\,dex in their Fig. 13). To determine if our simulated spectra are representative, we used their variance statistics for the COSMOS2020 $1.44\textrm{\,deg}^2$ and the anticipated E-COSMOS $4\textrm{\,deg}^2$ fields (Table \ref{table:GkogkouVariance}). While the [CII] sample variance of the power spectra components for COSMOS2020 is approximately $1.5$ times that of E-COSMOS, it is $<$0.25 for all cases excluding the supplemental bands (at $z_\textrm{[CII]}>10$). These variance estimates suggest that the COSMOS2020-based simulations are likely consistent with the wider E-COSMOS field, at least within order-of-magnitude fluctuations, although field-to-field variations may remain. 
The low variance statistic of $<$0.25 means that any problematic overmasking in part of a field would be reflected in the whole field. Therefore we must be careful of the CO $J$\,=\,5$-$4 bright mask at 410\,GHz for observations in E-COSMOS, as well as any cases found in Euclid for E-CDF-S.

\begingroup
\setlength{\tabcolsep}{3pt} 
\begin{table}
\caption{Field-to-field variance estimates for different power spectra components of COSMOS2020 and E-COSMOS fields, from \cite{Gkogkou_2023}
.}
\centering
\begin{tabular}{c c c c c}
\hline\hline
 $\nu$ (GHz)  & CO cluster & CO shot & [CII] cluster & [CII] shot  \\
 \hline
G23 1.44 deg$^2$\\
\hline
90 &0.0731& 0.0281& 1.90 &1.58\\
150 &0.0983&0.0404 &0.559&0.542\\
225 &0.124&0.0539  &0.211&0.231\\
280 &0.141& 0.0629& 0.125&0.146\\
350 &0.161&0.0737 &0.0732&0.0914\\
410 &0.201&0.0983  &0.0645&0.0844\\
\hline
G23 4 deg$^2$\\
\hline
90 &0.0501& 0.0205& 1.19&1.00\\
150 &0.0673&0.0294 &0.349&0.342\\
225 &0.0852&0.0392  &0.132&0.146\\
280 &0.0967& 0.0458& 0.0781&0.0922\\
350 &0.110&0.0537 &0.0457&0.0577\\
410 &0.137&0.0716  &0.0403&0.0533\\
\hline
\end{tabular}
\label{table:GkogkouVariance}
\end{table}
\endgroup

We considered including [OIII] signal to recover it by masking foreground CO/[CII] line emission. However, this was not possible (Appendix \ref{appendix: OIII signal}). Similarly, [CI] signal had minimal contributions in the foreground, and also was excluded. Future work shall investigate cross-correlations involving these lines, where they will be more relevant.
\subsection{Impacts of noise}\label{sec:discussion_masking+noise}

Our analysis finds that the median [CII] models do exceed the 2000hr white plus correlated noise sensitivity above $300$\,GHz for all $k$ modes, and that the lower limit models usually exceed it for $k$$>$0.5\,$\textrm{Mpc}^{-1}$ (Fig. \ref{fig:Pk_NoExtrap}). However, fainter models are more impacted by the suppression from the Lorentzian frequency profile, so their lower $k$ modes are eclipsed by the $1/f$ correlated component of noise (Appendix \ref{appendix: Lorentzian V Gaussian}). Regardless of model, this result is only achievable near the full EoR-Spec observing time, Prime-Cam requires more effective observing time to achieve a median $S/N$ ratio above 5 to ensure [CII] signal recovery above 300\,GHz (approximately factor 5, with some mode dependence). We find the best $S/N$ for the clustering signal, and so should be likely to constrain the contribution of dimmer galaxies to the luminosity function. 
Removing additional principal noise components via PCA could help, but has diminishing returns as PCA removes structural signal components. Following C24, noise is less problematic for [CII] below 300\,GHz, but CO dominates there anyway. 
However, total CO emission always exceeds predicted noise below $300$\,GHz, including for the extended lower frequency bands, with $S/N>10$. Consequently, early analysis should focus on analysing the low frequency CO signal, which in turn can be used to model contaminant removal to recover [CII]. In any case, early observations with EoR-Spec will more accurately characterise the instrumental and atmospheric systematics, thus allowing us to better model and remove them (Steve Choi priv. comm.). 
It will also identify what is required for next-generation instruments (such as spectrometers-on-chip) to improve LIM measurements.

\begin{figure}[t]
 \centering
 \includegraphics[width=\linewidth]{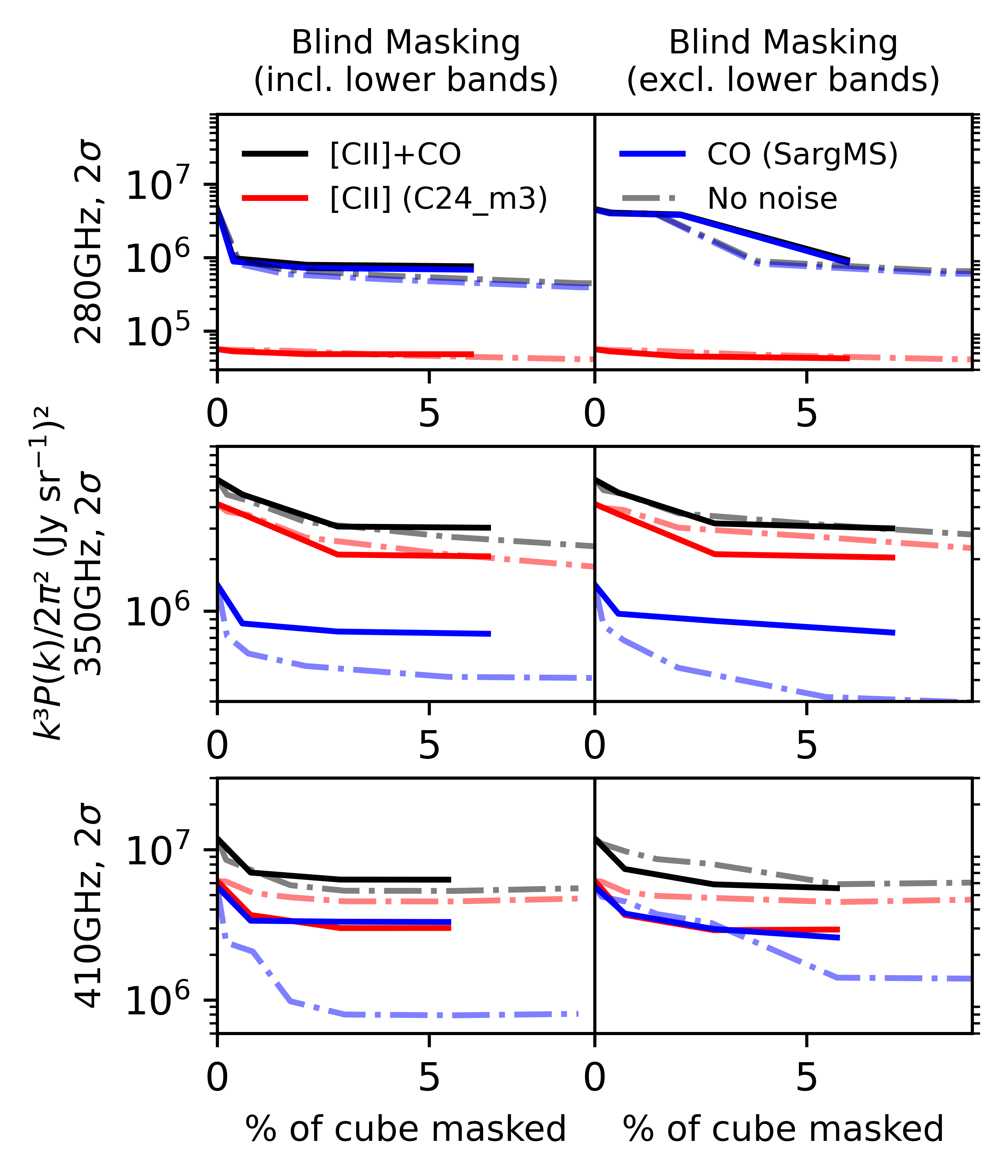} 
 \captionof{figure}{Demonstrating challenges in blind masking signal recovery including cleaned white and correlated noise in our cubes. The left and right subplots show blind masking including or excluding lower frequency bands respectively. All use 2$\sigma$ masks, with the upper subplots using 280\,GHz for the bright CO and [CII] models (SargMS, C24\_m3), middle suplots using 350\,GHz, and the lower at 410\,GHz with the same models. Noise prevents blind masking from working as effectively in the latter two cases by up to 0.5\,dex (compared ``no noise'' dashdot), and while it does not meaningfully impact 280\,GHz, the CO signal is too strong in that case.}
 \label{fig:RoyNoise}
\end{figure}

The magnitude of noise also impacts blind masking, which we tested by adding noise directly to those cubes. From our results in Fig. \ref{fig:RoyNoise}, the CO power spectra fails to be reduced by up to $\sim$0.5\,dex as many bright voxels are no longer from line emission, so the automated detection and pairing fail. Any implementation of blind masking therefore requires additional verification, such as by trying to fit spectral line profiles to the frequency dimension to distinguish between noise and galaxy emission. However, the noise has minimal impact on targeted masking, as the positions of masked voxels are not influenced by the noise of the map (thereby following the $S/N$ of Fig. \ref{fig:Pk_NoExtrap}). 

\subsection{Applications of additional techniques}\label{sec:discussion_improvements}

Overall, our extension of C24 to include CO based on empirical data indicates that CO and [CII] power spectra detections will be plausible for EoR-Spec when masking. However, early detections will be difficult considering noise and instrumental systematics of EoR-Spec, the latter of which can only be quantified after observations start, as seen with signal aliasing for COMAP (\citealt{Lamb_2022}, \citealt{Lunde_2024}) or shifts in kinetic inductance detector resonance frequencies for CONCERTO (\citealt{Bounmy_2022}, \citealt{Hu_2024}). In the interim we must consider statistics less sensitive to noise or contaminant line emission, including cross-spectra \citep{Roy_2024, Agrawal_2026}, line stacking (\citealt{Agrawal_2025}, \citealt{Dunne_2025}, \citealt{Moore_2025}), and spectral-line deconfusion (\citealt{Cheng_2020}, Oak in prep.).

LIM cross-correlation can recover information on the underlying mass density field (\citealt{Kovetz_2017}, \citealt{Dumitru_2019}). By taking the cross-spectra of two frequency channels containing line emission of the same galaxies, components not shared between the maps will fall out, mitigating the impact of noise and non-structured contaminants. Within EoR-Spec alone, cross-correlation can recover [CII] and [OIII] at 225 and 410\,GHz ($7<z<8$), as discussed by \cite{Roy_2024}. Additional frequency bands for the same field inherently make this technique stronger, such as with the EoR-Spec extension or other instruments. These include the balloon-based TIM  covering Euclid at 714$-$1250\,GHz ($0.5<z_\textrm{[CII]}<1$, \citealt{Vieira_2020}) and COMAP-Wide covering CO in wider areas than the Pathfinder (26$-$34\,GHz, \citealt{Cleary_2022})
. 
However, differing beam sizes and map coverage of these instruments will limit the accessible $k$ modes. In addition, ``triplets'' of cross-spectra at the same redshift can recover the auto-spectra via statistical arguments, becoming more feasible with additional frequency bands and so encouraging cross-instrument coverage (\citealt{Beane_2019} and \citealt{McBride_2024}). This will be the primary subject of our follow-up work.

Phase-space spectral-line deconfusion provides a different approach to the above techniques, and can avoid their inherent limitations. If multiple spectral lines originating from the same galaxy are observed, its redshift can be determined by fitting the emission to a set of templates \citep{Cheng_2020}. Then, by subtracting the CO line emission, the underlying [CII] signal can be extracted, even in bands with a low $S/N$. As with blind masking, template matching is less sensitive to catalogue incompleteness as it can fit to unknown repeating frequency structures.
For areas where catalogue data is more complete, this ancillary information can help deconfusion in extremely low $S/N$ regime, and so this technique can also be more robust against noise compared to blind masking.
Oak et al. (in prep.) will discuss its feasibility for a FYST-like deep sky survey. 

In addition, stacking emission from known bright sources for LIM can constrain low $k$ clustering signal even with high noise \citep{Dunne_2025, Dunne_2026}. Applying this technique to known clusters within the COSMOS field in the commissioning stages of EoR-Spec, such as AzTEC-3 and CRLE (\citealt{Riechers_2010}, \citealt{Pavesi_2018}), would allow for greater effective observation time without the need for extensive map scanning.

We also briefly investigated masking in the context of a successor to Prime-Cam, with up to thrice the resolution and a narrower frequency profile. We expect such an instrument to improve in masking effectiveness by up to a factor of two (Appendix \ref{appendix: nextgen}). Higher resolution instruments also allow more flexible masking strategies by adjusting mask width, which (depending on line model) may be sufficient to recover [CII] at 280\,GHz.

Finally, we anticipate that our techniques will be applicable to other fields, especially the Euclid deep fields which will be relevant to E-CDF-S and so EoR-Spec DSS. Consequently, we will apply our methods to future Euclid data releases.
\section{Conclusion} \label{sec:conclusion}

In this work, we extended the formulation of C24 for creating empirically-based LIM intensity cube simulations from existing catalogues in the frequency bands covered by EoR-Spec DSS. We produced cubes with multiple sources of line emission, primarily CO and [CII]. This was also done for two additional frequency bands proposed by \cite{Roy_2024} for a hypothetical next-generation instrument. We tested masking techniques within the E-COSMOS field, on both lower signal limits as well as estimates formed from extrapolated catalogues. This included targeted masking with source catalogues as done by \cite{VanCuyck_2023} and \cite{Karoumpis_2024}, and blind masking techniques which target bright voxels without using an external catalogue. We also compared these to early noise estimates for EoR-Spec. From our predicted simulated results we found:
\begin{itemize}
\item A SLED-based approach to CO line emission whilst using a $L_{\textrm{IR}}$-based bulk property model for CO $J$\,=\,1$-$0 shows that COSMOS2020 CO line luminosity functions lie within $1\sigma$ limits of dedicated CO galaxy studies using ALMA, the VLA and NOEMA
. These models are biased towards brighter emission at low redshift, and so provide a challenging case to recover [CII] via masking. CO and [CII] intensities and power spectra are always within 0.5\,dex above 300\,GHz, though CO dominates at lower frequencies.

\item Targeted masking successfully recovered [CII] signal above 300\,GHz when using wide masks ($\sigma$\,$>$\,1\,voxel), with $P_\textrm{[CII]}/P_\textrm{CO}\approx5$, but failed at lower frequencies. 
CO $J$\,=\,5$-$4 at 410\,GHz was the only CO transition where we observed overmasking for E-COSMOS, giving confidence for masking in that field. However, this only held when using a sufficiently complete foreground catalogue, otherwise key contaminant signal is missed. This requirement is met for COSMOS2025, with mass completeness gain of $>$1\,dex (90\% complete at $10^7\,M_\odot$ compared to COSMOS2020 at $10^8\,M_\odot$), but due to covering one-third of the area it reduces the accessible $k$ modes and results are more subject to field-to-field variance.

\item Blind masking supplements targeted masking by identifying and removing bright CO emission without relying directly on an external foreground catalogue, and can outperform it when the catalogue is complete. Extending EoR-Spec or doing cross-instrument science will improve the cleaning by $\sim$0.5\,dex (Fig. \ref{fig:Blindwithnolowfreq}). 

\item White and correlated noise within EoR-Spec is stronger at high frequency, compromising blind masking especially above 300\,GHz. Tentative auto power spectra detections should be possible with targeted masking as that is not directly affected by noise, but additional effective observing time is needed to achieve a $S/N>5$ (approximately factor 5 depending on $k$ mode, with clustering signal being more recoverable), though next generation on-chip spectrometers could mitigate these issues. For this reason, [CII] recovery depends on the underlying [CII] and CO emission models, else noise can complicate signal recovery if the foreground catalogue is incomplete. 
However, CO signal below 300\,GHz should be recoverable under the conditions explored here.
\end{itemize}
\noindent In this way, our study builds on previous empirically motivated [CII] LIM forecasts, and demonstrates that contaminant mitigation using comparatively direct approaches such as targeted and blind masking may be feasible for the EoR-Spec DSS. It also provides a framework for investigating more advanced cross-instrument analyses, such as cross-correlation, in future work.

\begin{acknowledgements}
We wish to thank members of the CCAT Collaboration (including Nick Battaglia, Dongwoo Chung, Thomas Nikola and Steve Choi) as well as Himadri Saha, Yun-Ting Cheng, Akira Endo, Elena Marcuzzo, Anirban Roy, Teo Topkaras, and Mathilde Van Cuyck for their valuable input. 

We wish to acknowledge the contributions of the individual components of the COSMOS 2020 survey (\href{https://cosmos.astro.caltech.edu/}{https://cosmos.astro.caltech.edu/}). In specific, this work is
based on observations collected at the European Southern Observatory under ESO programme ID 179.A-2005 and on data products produced by CALET and the Cambridge Astronomy Survey Unit on behalf of the UltraVISTA consortium. It is also based on data products from observations made with ESO Telescopes at the La Silla Paranal Observatory under ESO programme ID 179.A-2005 and on data products produced by CALET and the Cambridge Astronomy Survey Unit on behalf of the UltraVISTA consortium.
The CCAT project, FYST and Prime-Cam instrument have been supported by generous contributions from the Fred M. Young, Jr. Charitable Trust, Cornell University, and the Canada Foundation for Innovation and the Provinces of Ontario, Alberta, and British Columbia. The construction of the FYST telescope was supported by the Gro{\ss}ger{\"a}te-Programm of the German Science Foundation (Deutsche Forschungsgemeinschaft, DFG) under grant INST 216/733-1 FUGG, as well as funding from Universit{\"a}t zu K{\"o}ln, Universit{\"a}t Bonn and the Max Planck Institut f{\"u}r Astrophysik, Garching. The construction of EoR-Spec is supported by NSF grant AST-2009767. The construction of the 350\,GHz instrument module for Prime-Cam is supported by NSF grant AST-2117631.
In addition, this is based on observations collected at the European Southern Observatory under ESO programme ID 179.A-2005 and on data products produced by CALET and the Cambridge Astronomy Survey Unit on behalf of the UltraVISTA consortium.

We gratefully acknowledge support provided by the Collaborative Research Center 1601 (SFB 1601 sub-projects C3, C6) funded by the Deutsche Forschungsgemeinschaft (DFG, German Research Foundation) – grant number 500700252. 

This work was supported by JSPS KAKENHI Grant Numbers JP23K20035, JP24H00004, JP26KJ0864, and JST SPRING Grant Number JPMJSP2108. This research was also supported by the International Graduate Program for Excellence in Earth-Space Science (IGPEES), a World-leading Innovative Graduate Study (WINGS) Program, the University of Tokyo.

We made use of the Python packages astroPy \citep{Astropy_Collaboration_2013,Astropy_Collaboration_2018,Astropy_Collaboration_2022}, numPy \citep{harris2020array}, scipy \citep{Virtanen_2020}, and matplotlib \citep{Hunter_2007}.
Many thanks to Dr. David Nichols for his article on use of colours for the colourblind: \href{https://davidmathlogic.com/colorblind/}{https://davidmathlogic.com/colorblind/}. In addition, we made extensive use of NASA's Astrophysics Data System Bibliographic Services throughout this work.

\end{acknowledgements}

\bibliographystyle{aa}
\bibliography{sample631}

\appendix
\section{Spectral line energy distribution}\label{appendix: SLED}
\begin{figure}[t]
 \centering
 \includegraphics[width=\linewidth]{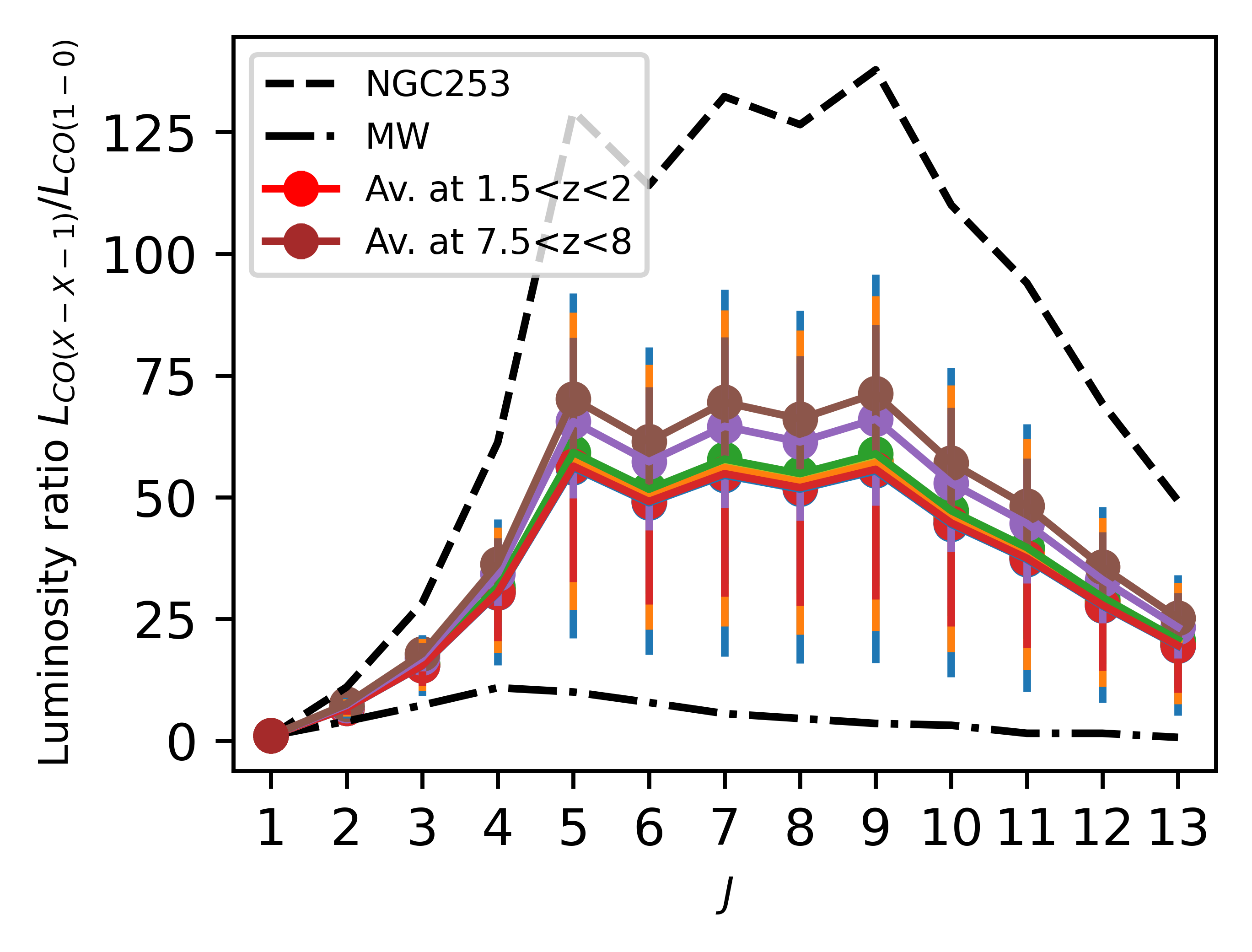}
 \captionof{figure}{Spectral line energy distributions in the form of CO luminosity ratios. We include NGC253, the Milky Way, and the medians of FARMER LP with $1\sigma$ error. FARMER LP SLED templates are subdivided in $\Delta z$\,=\,$0.5$ intervals, showing minimal difference with redshift.}
 \label{fig:SLED}
\end{figure}

Following \cite{Karoumpis_2024}, we used Spectral Line Energy Distribution (SLED) templates to propagate CO luminosity models past $J$\,=\,1$-$0 transition. We do this via Eq. \ref{eq:SLEDEQ}, taking coefficients from NGC253 and Milky Way data found in Table \ref{table:SLED}. Its free parameter $\mu$ varies from galaxy to galaxy, depending on its deviation from the galaxy main sequence for the given $\Delta z=0.5$ interval (Eq. \ref{eq:mu}). These SLED templates are represented visually in Fig. \ref{fig:SLED}, which shows their mean and standard variation. They do not meaningfully vary with redshift.
\begin{align}
\log_{10} \frac{L_{\textrm{CO\,}J=X-(X-1)}}{L_\odot}=((1-\mu)\textrm{SLED}_{\textrm{MW}(X-X-1)}\nonumber\\+\mu\textrm{SLED}_{\textrm{NGC253}(X-X-1)})\log_{10}\frac{L_{\textrm{CO\,}J=1-0}}{L_\odot},
\label{eq:SLEDEQ}
\end{align}
\begin{gather}
\mu=\frac{\Delta\textrm{Main Sequence}+1}{2.5}
\label{eq:mu}
\end{gather}
\begingroup
\setlength{\tabcolsep}{3pt} 
\begin{table}
\caption{CO SLED template parameters, used in Eq. \ref{eq:SLEDEQ}.}
\centering
\begin{tabular}{c c c}
\hline\hline
 Line Ratio (to CO $J$\,=\,1$-$0) & Milky Way (1) (3) & NGC253 (2) \\
\hline
CO $J$\,=\,1$-$0 & 1 & 1 \\
CO $J$\,=\,2$-$1 & 4 & 11 \\
CO $J$\,=\,3$-$2 & 7.29 & 28.25\\
CO $J$\,=\,4$-$3 & 10.88& 61.25\\
CO $J$\,=\,5$-$4 & 10 & 125.25\\
CO $J$\,=\,6$-$5 & 7.85& 114\\
CO $J$\,=\,7$-$6 & 5.57& 132.25\\
CO $J$\,=\,8$-$7 & 4.56& 126.5\\
CO $J$\,=\,9$-$8 & 3.54& 137.75\\
CO $J$\,=\,10$-$9 & 3.17 & 110\\
CO $J$\,=\,11$-$10 & 1.52& 94\\
CO $J$\,=\,12$-$11 & 1.52 & 69.25\\
CO $J$\,=\,13$-$12 & 0.683& 49.25\\
\hline
\end{tabular}
\tablebib{(1) \cite{Carilli_2013}, (2) \cite{Mashian_2015}, (3) \cite{Wilson_2017}}
\label{table:SLED}
\end{table}
\endgroup
\section{Low redshift discrepancies}\label{appendix: low z}

The most crucial assumption we made is that the assigned photometric galaxy properties from FARMER LP are accurate, otherwise bulk property models give incorrect luminosities and skew the resulting power spectra. Extrapolation would compound any existing issues by duplicating erroneous galaxies. There is evidence that such misassignments can occur, as shown when machine-learning techniques were applied to COSMOS2020 \citep{Asadi_2025}. This work implied that the photometric code over-assigned galaxies as star-forming instead of quiescent at $0.2<z<3.5$ (20\% instead of 10\% at low $z$), thus skewing SFRs and stellar masses.

\begin{figure}[t]
 \centering
 \includegraphics[width=\linewidth]{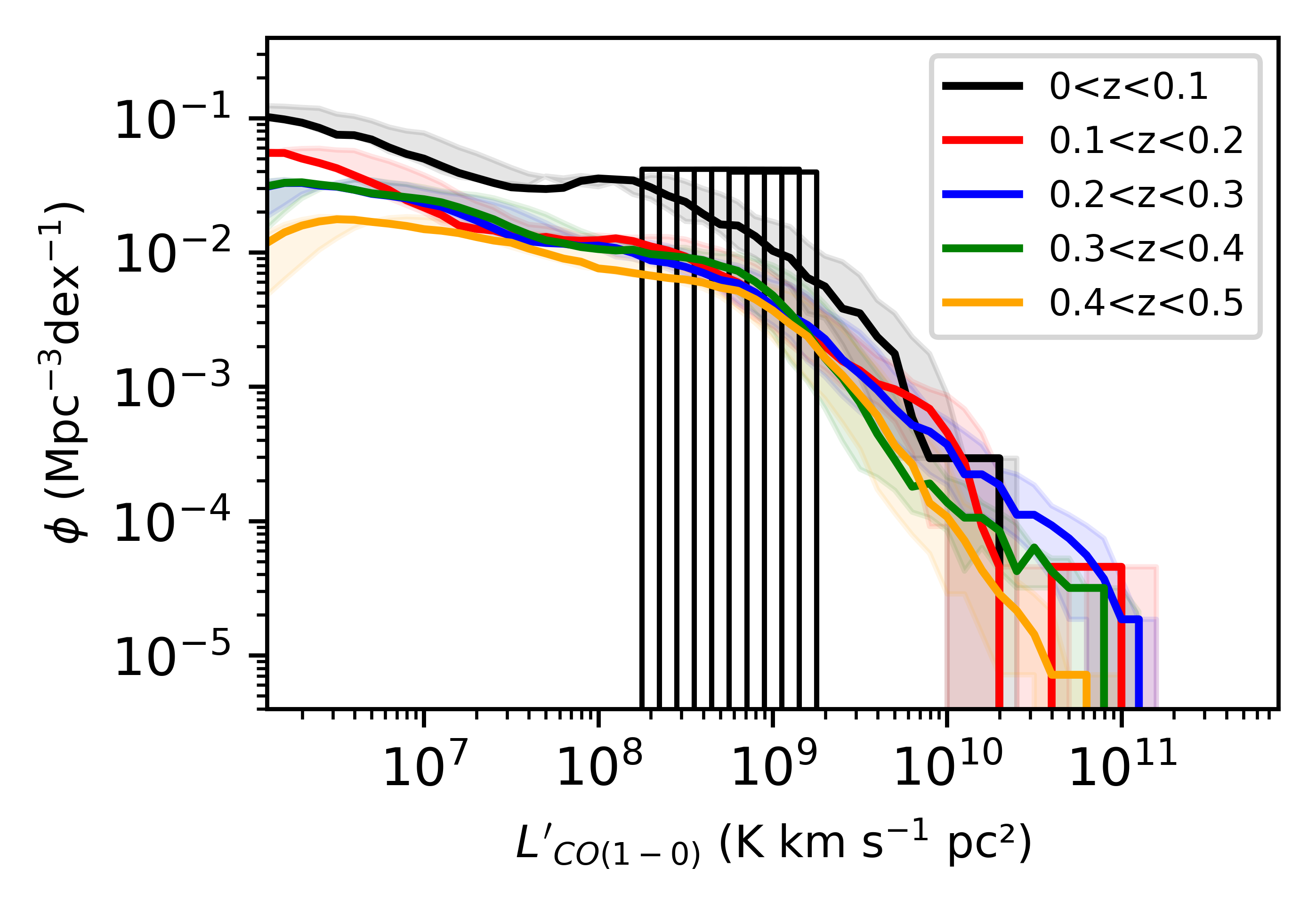}
 \captionof{figure}{CO $J$\,=\,1$-$0 line luminosity functions of FARMER LP at $z<0.5$, in intervals $\Delta z=0.1$, with the ASPECS luminosity function of $z\approx0.153$ overlaid. We see a disproportionate number of sources below $z<0.1$, providing a greater contribution to CO luminosity.}
 \label{fig:LowZLumFunc}
\end{figure}

While we anticipated this for sources past $z>6$ which have lower $S/N$ and sample numbers (see C24), when making early maps we found that problems within FARMER LP are most impactful at low redshift (Fig \ref{fig:LowZLumFunc}). 
Here, we found a disproportionately high number of low luminosity galaxies at $z<0.1$, $\sim$0.5\,dex higher than high redshift bands, with a total intensity that skewed the power spectra up. While these sources contribute less than higher luminosities via log-law arguments with the Schechter curves, the sheer number is non-negligible. We also found similar relations in the SFR functions, and found far more galaxies with low mass in the same range (which would skew the galactic main sequence parameter to calibrate the SLED templates, Eq. \ref{eq:mu}). These low $z$ errors are supported by the CANDELS ratio being low for $z<0.5$ (Table \ref{table:CANDELSlims}), as well as \cite{Weaver_2022} and \cite{Weaver_2023} showing a decrease in the galaxy stellar mass distribution (in Figs. 14, 20 and Fig. 2 respectively). 

\begin{figure}[t]
 \centering
 \includegraphics[width=\linewidth]{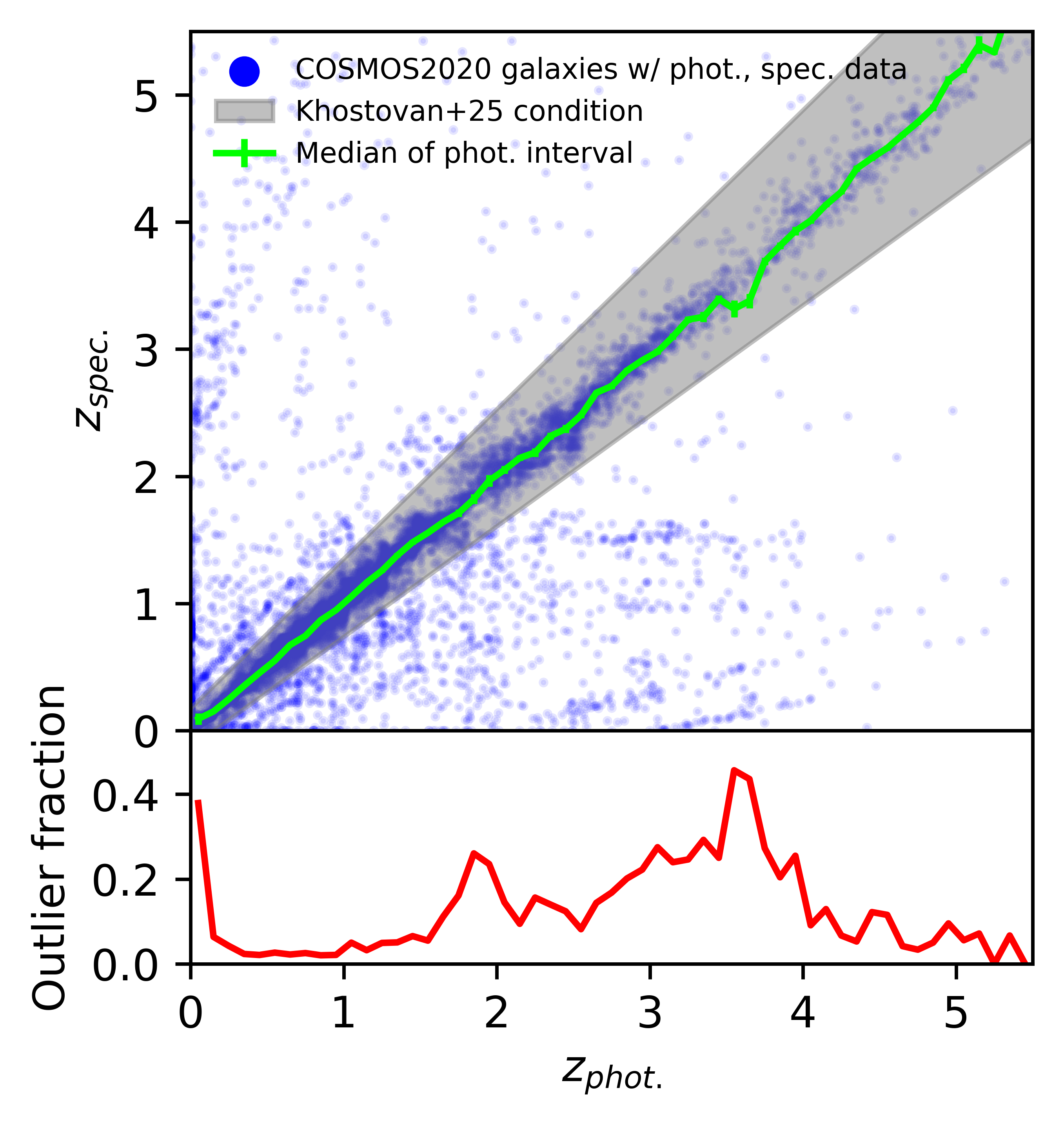}
 \captionof{figure}{Comparing photometric and spectroscopic redshifts in COSMOS2020 using \cite{Khostovan_2025}. The upper subplot shows this with the grey shaded region indicating the ``valid'' region (as defined in their work). The lower subplot indicates the fraction of galaxies that do not lie within this region for photometric $z$. This is poor around cosmic noon and $z<0.1$. The former is expected due to how automated fitting codes can have trouble in that region (see Fig. 17 from \citealt{Weaver_2022}), but due to the narrower relative spread this is unlikely to be troublesome. However, this is more problematic for the latter case, because of how high spectroscopic $z$ galaxies are misconstrued as low $z$.}
\label{fig:KhostovanComp}
\end{figure}

A potential cause was shown by \cite{Khostovan_2025}, where they found a greater deviation between photometric and spectroscopic redshifts at $z<0.1$ (their Fig. 11, our Fig. \ref{fig:KhostovanComp}). This different to the deviation at $1.5z<3.5$, explained by \cite{Weaver_2022} as the Balmer break being redshifted out of their observed frequency range (their Fig. 17). In this way, it is possible that galaxies at $z<0.1$ were mis-assigned to high redshift, and vice versa, with corresponding bulk property errors. 

Correspondingly, making CO maps for $z<0.1$ is less reliable, and so we do not include any of those galaxies in our cubes. In our case, this meant not including CO $J$\,=\,3$-$2 emission for 350\,GHz and CO $J$\,=\,2$-$1 emission for 225\,GHz. The adjusted atmospheric transmission window meant that we avoided the lowest $z$ signal for the CO $J$\,=\,4$-$3 line. While this is vexing due to the importance of low CO transition signal, it is safe to assume that in observations we will be able to identify these local galaxies more easily and mask them out manually. 

\section{Lorentzian frequency profile}\label{appendix: Lorentzian V Gaussian}

\begin{figure}
 \centering
 \includegraphics[width=\linewidth]{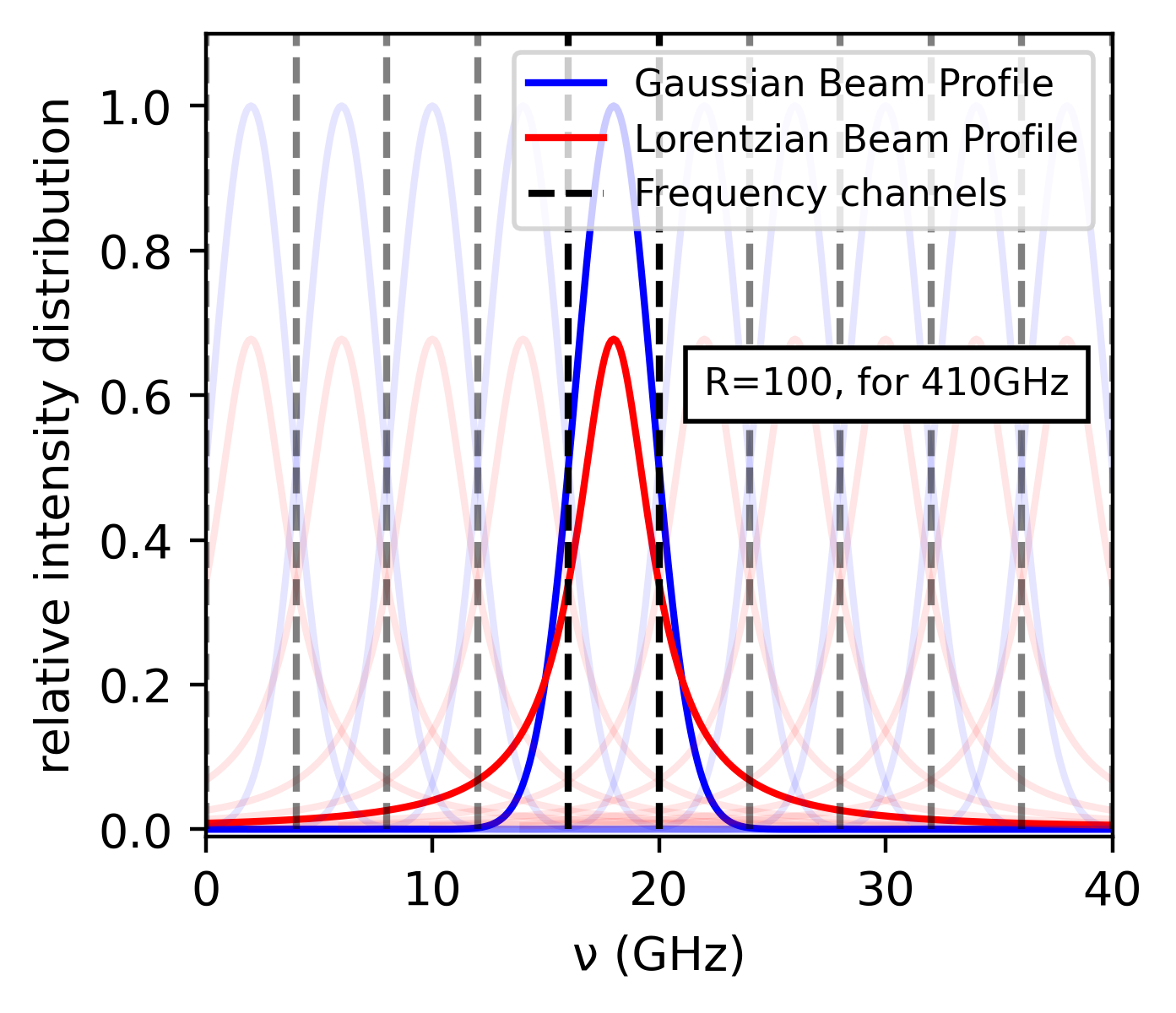}
 \captionof{figure}{Relative intensity distribution over the frequency channels for a Lorentzian and Gaussian profile at 410\,GHz (red and blue respectively), with the Lorentzian having greater spread.}
 \label{fig:LorVGaussBeam}
\end{figure}

\begin{figure}
 \centering
 \includegraphics[width=\linewidth]{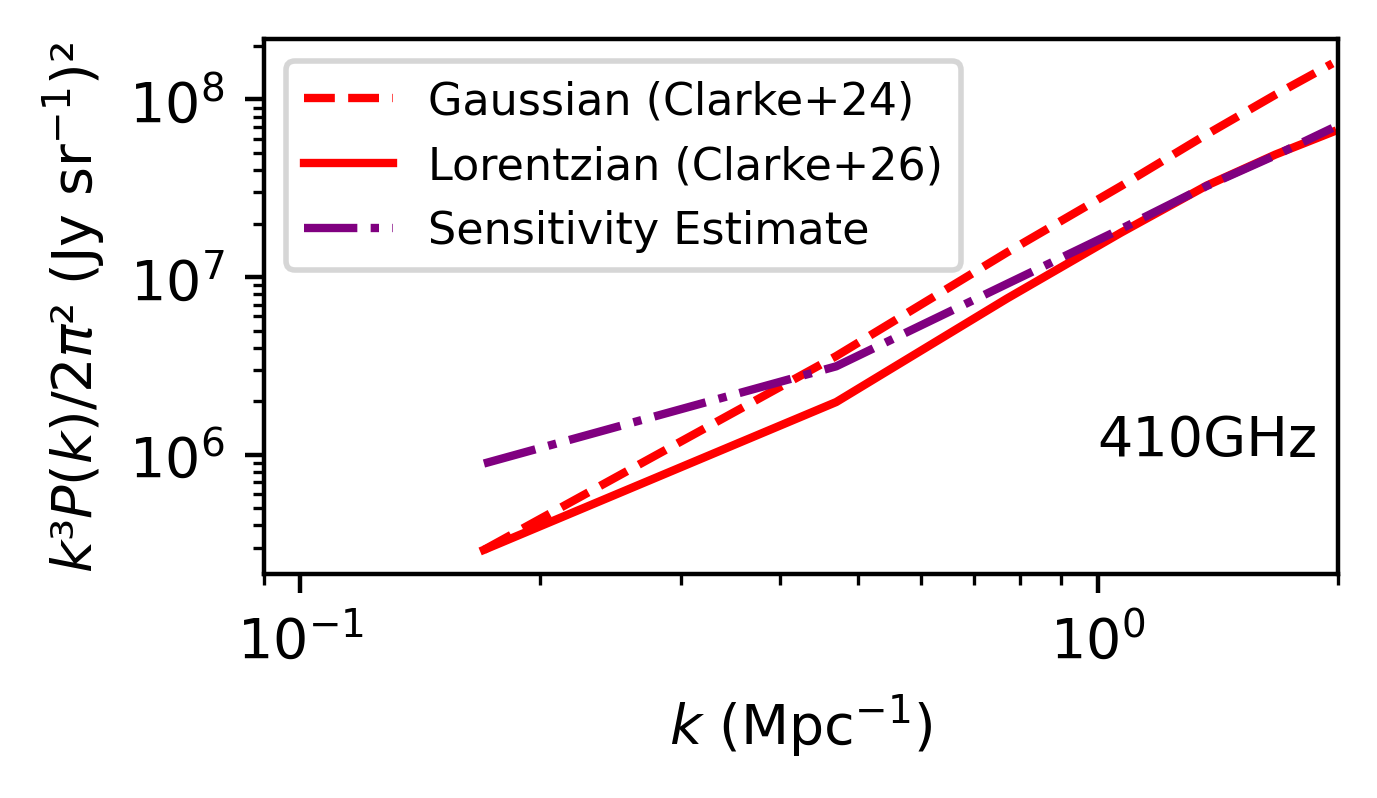}
 \captionof{figure}{Comparing the power spectra of this work and C24 (red solid vs. dashed): Lorentzian vs. Gaussian frequency profile, with and without subgridding. Due to this, our shot-noise is reduced by up to $\sim$0.3\,dex. Sensitivity limits (purple dash-dot) are shown for contrast.}
 \label{fig:LorVGaussSpectra}
\end{figure}

Whilst earlier works assumed that the frequency profile of EoR-Spec could be characterised by a Gaussian shape (\citealt{Chung_2020}, \citealt{Karoumpis_2021,Karoumpis_2024}, C24), in actuality it better resembles a Lorentzian, with the respective shapes found by Eqs. \ref{eq:Gauss1D}, \ref{eq:Lorent1D}, where $\Gamma=\sigma\sqrt{2\ln{2}}$ and the FWHM of the profile is treated as one frequency channel. In all cases, the EoR-Spec beam in map space can be described by a 2D Gaussian (Eq. \ref{eq:2DGaussEq}). These are parameterised as:

\begin{align}
G(\mu_\nu,\nu,\sigma)=\frac{1}{\sigma \sqrt{2\pi}}\exp{\left (-\frac{(\mu_\nu-\nu)^2}{2\sigma^2}\right)},
\label{eq:Gauss1D}
\end{align}

\begin{align}
L(\mu_\nu,\nu,\Gamma)=\frac{1}{\pi}\frac{\Gamma}{(\mu_\nu-\nu)^2+\Gamma^{2n}},
\label{eq:Lorent1D}
\end{align}

\begin{align}
G(\mu_\textrm{RA},\mu_\textrm{Dec},\textrm{RA},\textrm{Dec},\sigma)=\frac{1}{2\pi \sigma^2}\textrm{exp}\bigg( -\frac{(\mu_\textrm{RA}-\textrm{RA})^2}{2\sigma^2}\nonumber\\-\frac{(\mu_\textrm{Dec}-\textrm{Dec})^2}{2\sigma^2}\bigg),
\label{eq:2DGaussEq}
\end{align}

\noindent where $\mu_\textrm{RA}$, $\mu_\textrm{Dec}$, and $\mu_\nu$ are the locations of the galaxy in question, and $n$ is the order of the Lorentzian profile (by default, $n=1$). As per our cube dimensions the FWHM of the frequency profile is $3$ voxels, so the standard deviation $\sigma=3/(2\sqrt{2\ln2})$ voxels and half-width half maximum $\Gamma=3/2$ voxels. Correspondingly, we find that the Lorentzian profile exhibits a greater spread in frequency space, as exhibited in Fig. \ref{fig:LorVGaussBeam}, where $76\%$ vs. $53\%$ lies within 1 frequency channel. As noted by \cite{Marcuzzo_2025}, this reduces the contrast between voxels, which (combined with our subgridding into $3\times3\times3$ voxels) can dampen the shot-noise by up to 0.2\,dex (see Fig. \ref{fig:LorVGaussSpectra}
). While this decrease in shot-noise is low, we must consider it when attempting to recover signal, or when by spectral fitting.

\section{COSMOS2025}\label{appendix: COSMOS2025Dev}

\begin{figure*}
 \centering
 \includegraphics[width=\linewidth]{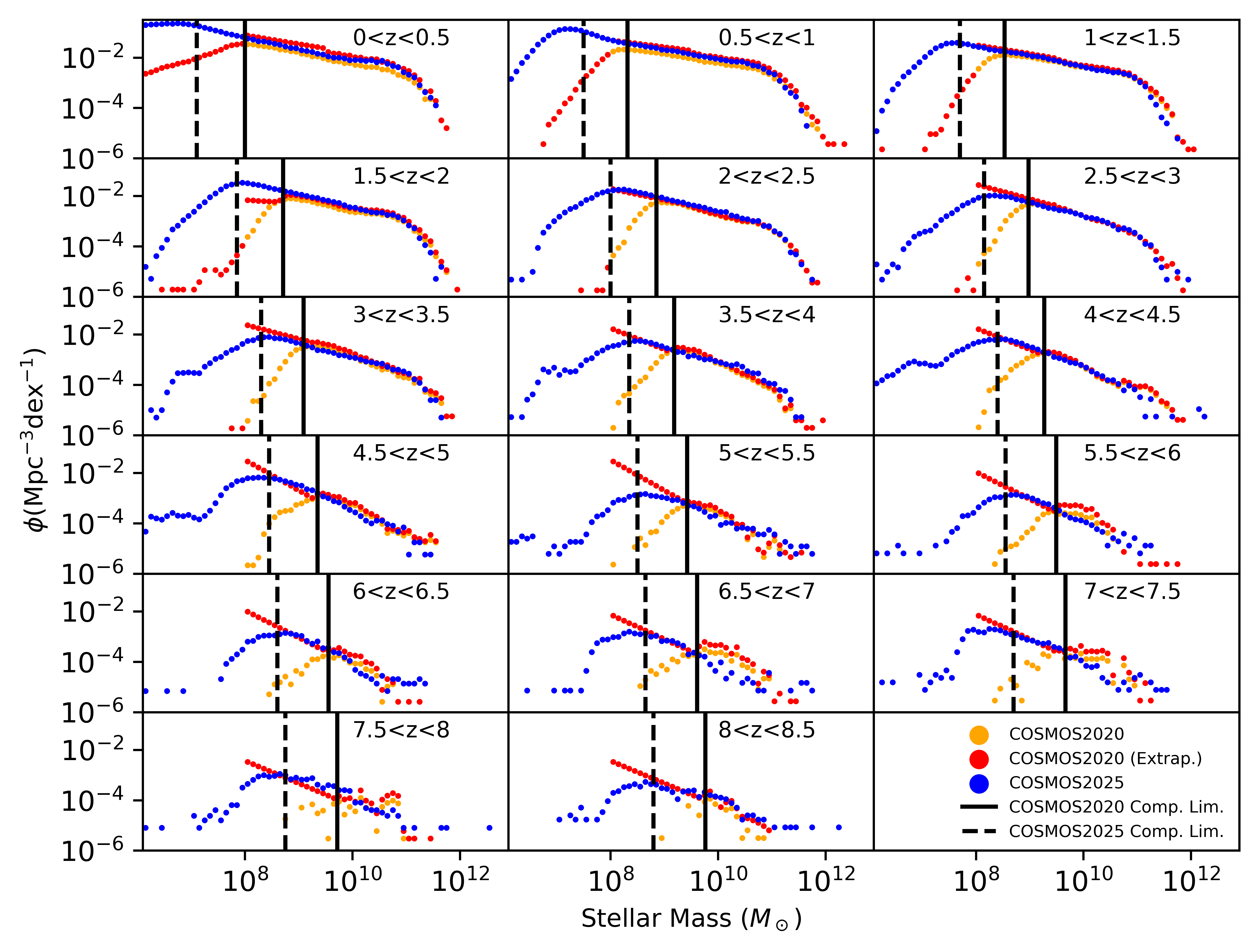}
 \captionof{figure}{Comparing the mass functions of FARMER LP and COSMOS2025 for each $\Delta z=0.5$ interval. Orange indicates COSMOS2020 without extrapolation, red with extrapolation, blue indicates COSMOS2025, and the solid and dashed vertical lines indicate COSMOS2020/5 completeness limits. Here, COSMOS2025 aligns with the extrapolated case of FARMER LP. There is greater deviation at $z>4.5$ due to increased uncertainties, especially below the mass completeness limits.}
 \label{fig:COSMOS2025_example}
\end{figure*}

The JWST COSMOS-Web program \citep{Casey_2023} observed a $0.54\textrm{\,deg}^2$ region of the COSMOS field, with JWST providing near-IR imaging, resulting in the release of COSMOS2025 \citep{Shuntov_2025}. This catalogue has reduced redshift uncertainties and improved mass completeness over all redshift ranges, corresponding to $\sim$1\,dex across all redshifts. For example, at $z\approx0.25$ the 90\% Mass completeness limit of COSMOS2020 is $10^8 M_\odot$, but is closer to $10^7 M_\odot$ for COSMOS2025. While the smaller field (one-third the area of COSMOS2020) makes COSMOS2025 less suitable to use for LIM forecasts, it can be used to calibrate our assumed extrapolation metrics, as shown by comparing the mass functions of COSMOS2025 and FARMER LP (Fig. \ref{fig:COSMOS2025_example}).

For most redshifts, the mass function of COSMOS2025 aligns with our extrapolated sample. Considering its greater completeness this shows that extrapolation is an appropriate first-order approximation to the cosmic field. COSMOS2025 is also able to probe deeper into the mass function, below our assumed cut-off of $10^8 M_\odot$. We still choose not to extend our method include these galaxies as low mass galaxies produce proportionally less shot-noise compared to high mass galaxies so losing them is acceptable \citep{VanCuyck_2023}, and our understanding of them is limited. We did find greater deviation at $z>4.5$, where our sample has a greater $\Phi$ below the COSMOS2025 mass limit. Considering this is where COSMOS2025 is incomplete, discrepancies are expected. This is also seen for $z>6.5$ at the high end of the mass function, concordant with COSMOS2020 being limited to the ultra-deep slices of UltraVISTA. Nevertheless, COSMOS2025 is a reassuring check of extrapolation from FARMER LP.
\section{Uncertainty in galaxy redshift}\label{appendix: RandomLocMasking}

Uncertainties in photometric redshift are more concerning than bulk property errors as they govern the frequency location of galaxies. These errors only affect shot-noise by up to 0.1\,dex as discussed in the appendix of C24, but are more problematic for targeted masking. Our analysis assumed that the foreground catalogue locations perfectly align with the intensity peaks in the cubes, ensuring CO signal is fully masked out. Consequently, redshift/frequency errors (up to 15\% for some galaxies) could result in missing CO emission or accidental overmasking.

\begin{figure}
 \centering
 \includegraphics[width=\linewidth]{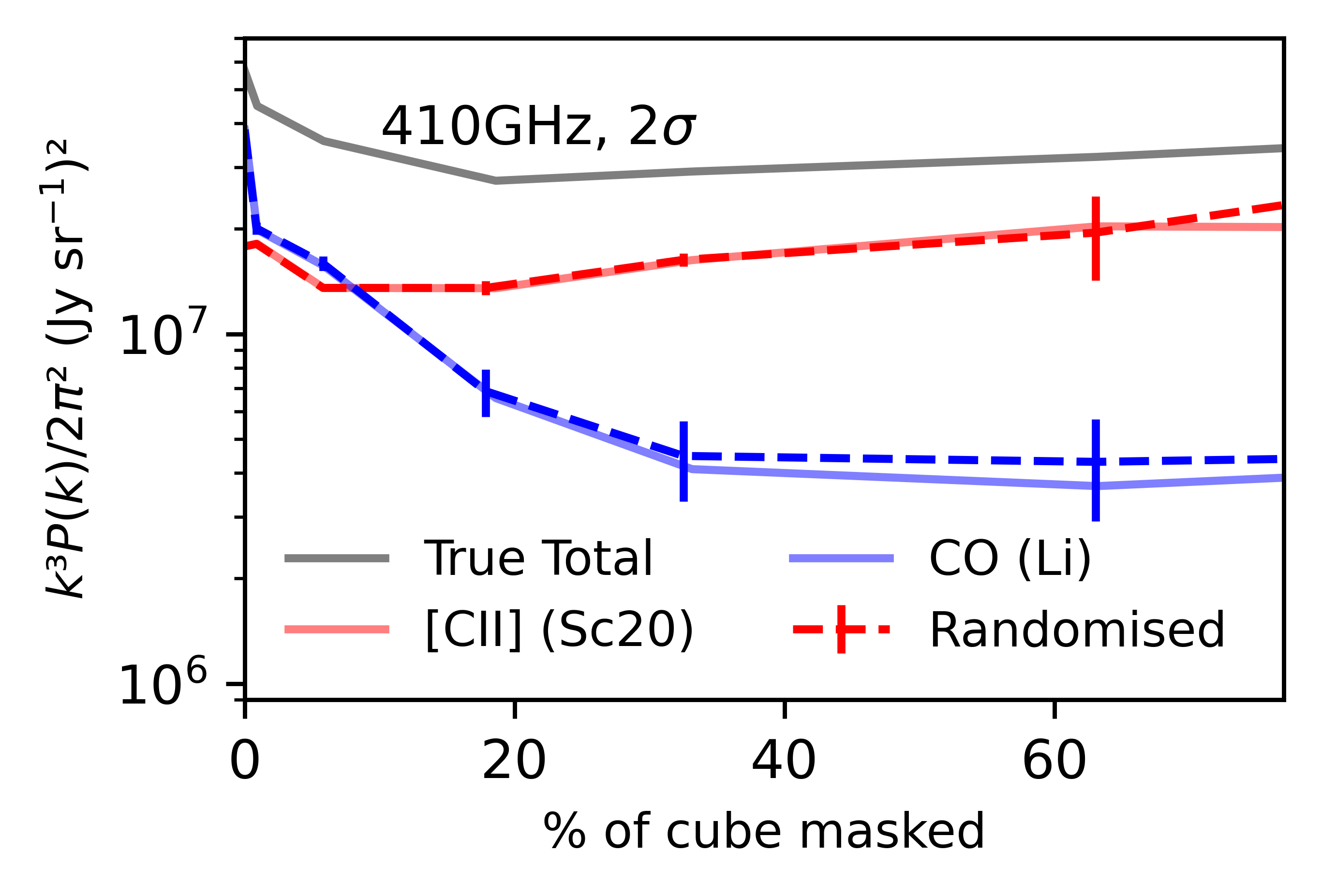}
 \captionof{figure}{Demonstrating the impact of redshift uncertainty on a given intensity cube. For 410\,GHz, SargSB CO, Sc20 [CII], and combined signal with $2\sigma$ masking, we overlay the non-randomised case to ten realisations of the randomised case (dashed) with the corresponding 3$\sigma$ errors. These show that our simulations will likely be consistent with observations, though we do note that the error bars approach 0.2\,dex at higher masking volumes. }
 \label{fig:UncertaintyZ_showcase}
\end{figure}

To examine the importance of location uncertainties in masking, we took an intensity cube and reviewed the corresponding masks. For our example in Fig. \ref{fig:UncertaintyZ_showcase}, we used the FARMER LP standard deviation of all foreground galaxy redshifts to calculate the deviation in frequency space, and formed a photometric redshift probability distribution function (PDF) of Gaussian form for the mask frequency locations. We then made ten separate realisations, creating a new set of masks using the randomised locations.  
In our example, the effectiveness of CO cleaning was not meaningfully reduced, with the actual spectra lying cleanly within the 3$\sigma$ confidence limits provided by the randomised case. 
However, in our case, over $70\%$ of galaxies under $z<2$ had frequency deviation less than 3 cube slices, an ideal scenario. We were also using wide $2\sigma$ width masks, allowing effective masking even with a slight offset in frequency position. We anticipate this method would be less effective for samples with higher redshift uncertainties. Therefore, it will be important to create contaminant line emission PDFs for multiple realisations when doing foreground catalogue masking to test uncertainties.

We note that FARMER LP map locations have much lower uncertainties. 
In our sample, we found a maximum error of 5 voxels, with most galaxies showing a deviation of no more than 1 voxel. Any impacts of this on masking are negligible compared to the frequency spread.
\section{[OIII] signal}\label{appendix: OIII signal}

\begin{figure}
 \centering
 \includegraphics[width=\linewidth]{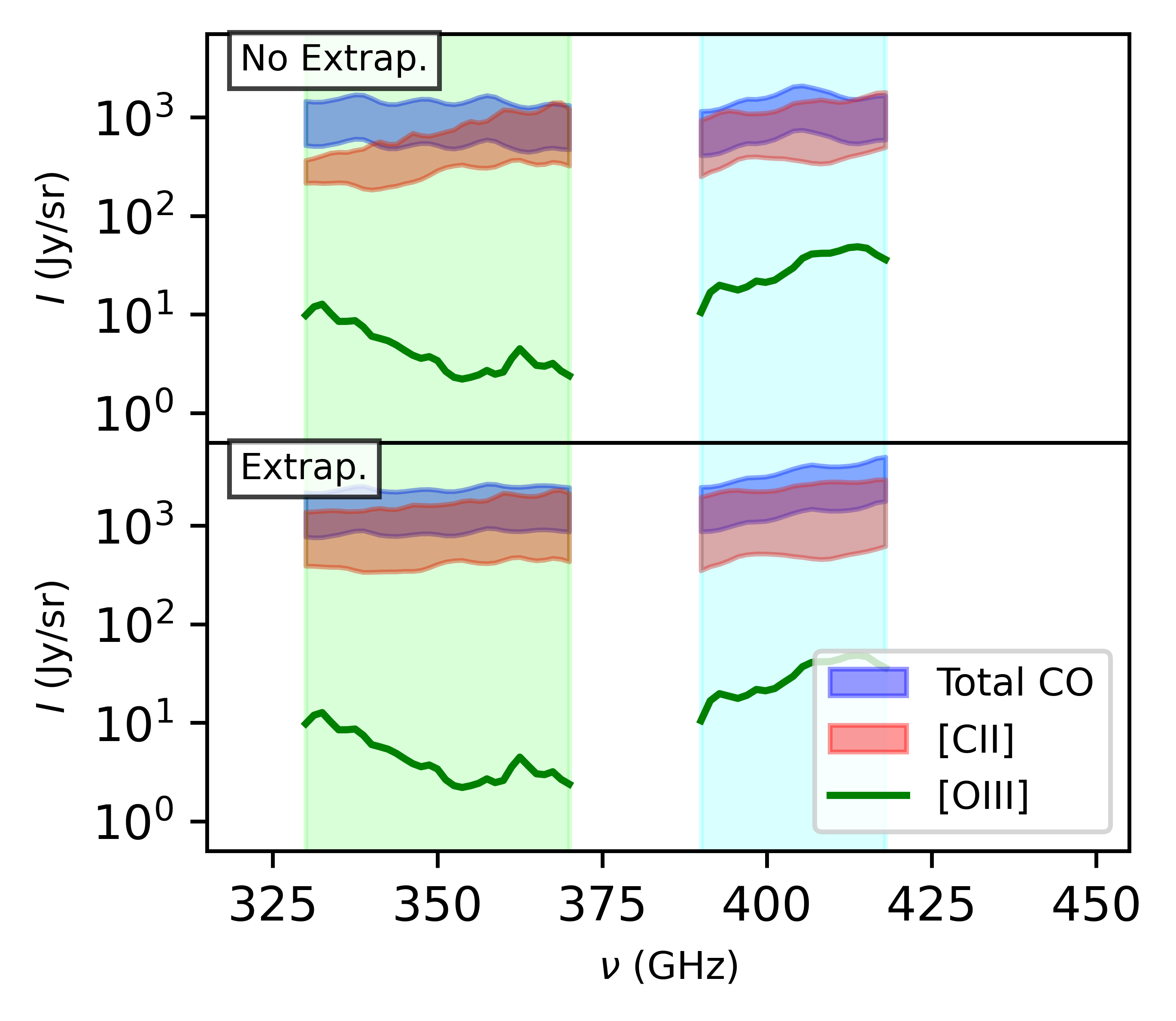}
 \captionof{figure}{As Fig. \ref{fig:IvsFreq}, but also including [OIII] in green. [OIII] emission is $\sim$$2\,\textrm{dex}$ dimmer, though due to being heavily reliant on individual bright sources we do not see the consistent steady emission decrease as in CO and [CII].}
 \label{fig:IvsFreq_OIII}
\end{figure}

Our analysis in this work focused primarily on CO, but we also made preliminary investigations into [OIII] emission as it will be crucial for cross-correlations within FYST and TIM. Due to the higher rest frequency of $3393\,\textrm{GHz}$, only the 410 and 350\,GHz EoR-Spec bands include [OIII] ($7.12<z<7.7$, $8.17<z<9.28$ respectively).

For $L_{\textrm{[OIII]}}$, we used values determined by the COSMOS2020 EAZY photometry code. We take this estimate because bulk property models for this line have not been calibrated at high $z$. When it came to extrapolated samples, we treated it similar to $L_{\textrm{IR}}$, that is we generated the $L_{\textrm{[OIII]}}$ values used in line modelling, for bias reasons indicated in Section \ref{sec:methodTomog}. As with $L_{\textrm{IR}}$, we used empirical SFR fits over mass-complete data in FARMER LP:
\begin{align}
\log_{10} \frac{L_{\textrm{[OIII]}}}{L_\odot}=7.41+0.672\log_{10}\frac{\textrm{SFR}}{M_\odot \textrm{yr}^{-1}},
\label{eq:OIII-SFR}
\end{align}

From our initial investigation of this line signal, we attempted to recover [OIII] by masking CO and [CII] signal. However, as shown in Fig. \ref{fig:IvsFreq_OIII}, [OIII] is approximately 2 orders of magnitude fainter than the other signal, so such recovery is impossible. By contrast, our initial cross-correlations show that [CII]$\times$[OIII] for $225/410\,\textrm{GHz}$ at $z\approx7.5$ will be detectable, so we save further analysis of [OIII] for future work.
\section{Next-generation instrument tests}\label{appendix: nextgen}

Prime-Cam, with its Lorentzian frequency profile and resolution R\,$=100$, will be important in developing submm LIM techniques. However, follow-up successor instruments are already in ideation and early development, planning to use the early findings of Prime-Cam to determine what is needed to improve LIM observations. To simplify these design decisions, we will consider cases with higher R and a narrower frequency profile. While resolving individual sources would still be impossible, these improvements would allow for greater distinction between sources at different redshift intervals, making structural features clearer. We expect this to have multiple benefits for LIM analysis, including for masking because of the relative ease in cleaning more distinct CO sources. These power spectra will have higher magnitudes as bright sources are spread over fewer voxels, in a similar manner to that described in Appendix \ref{appendix: Lorentzian V Gaussian}.

For our testing, we remade intensity cubes for the median CO and [CII] models, with the only differences being using $n=2$ for Eq. \ref{eq:Lorent1D} (giving a beam profile closer to a Gaussian) and changing the frequency bin width to reflect R\,$=200, \,300$. We then applied the standard masking methodologies, using masks of radius $1.5\sigma$. By default we used the ``new'' frequency channels to determine the size of masks in frequency space (that is, shrinking the effective frequency coverage). However, we also investigated applying masks which cover the same frequency range as R\,$=100$, effectively masking out more channels in the R\,$=200, \,300$ cubes. 

\begin{figure}
    \centering
 \includegraphics[width=\linewidth]{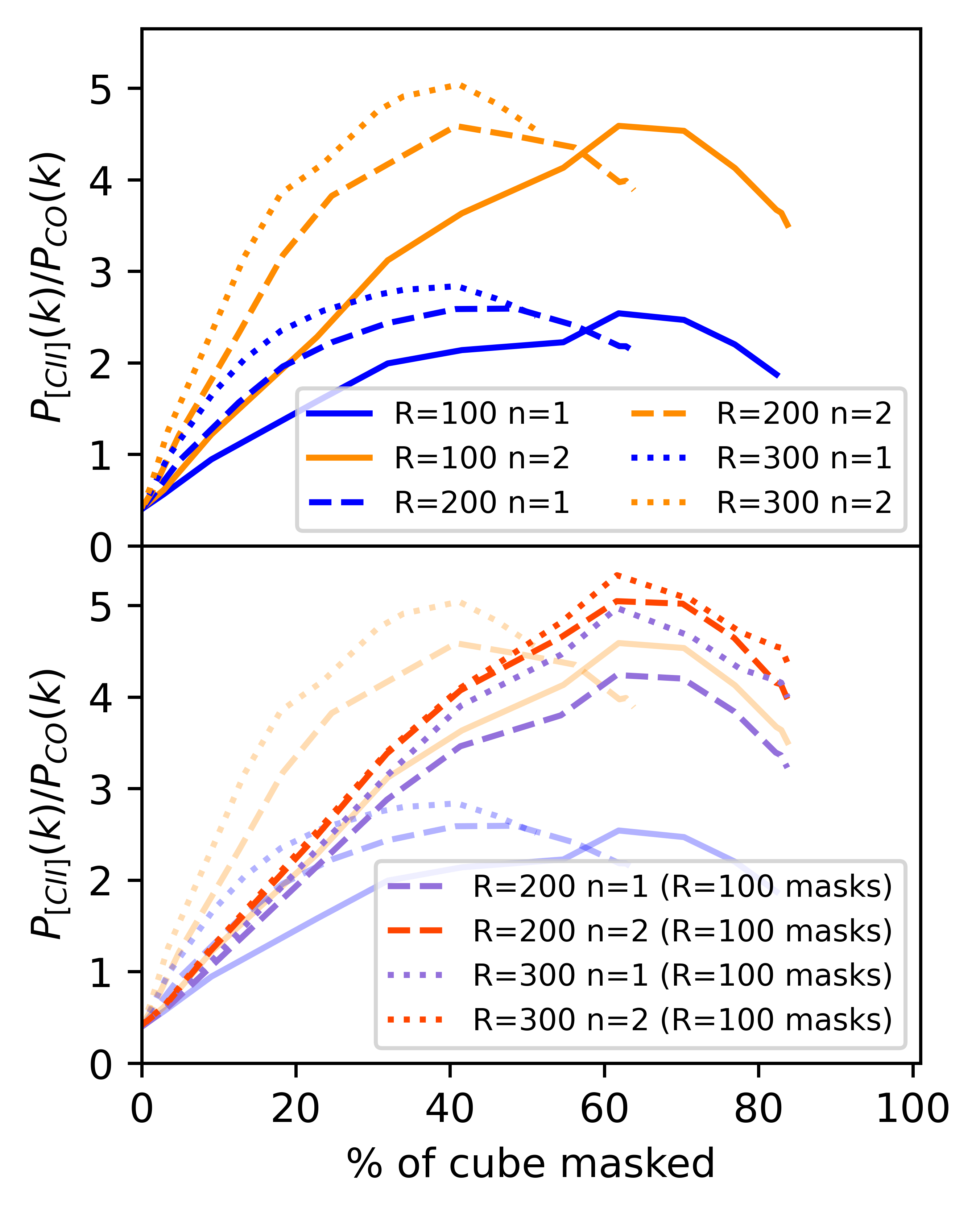}
 \captionof{figure}{Comparing masking efficiency, parametrised as the ratio of the [CII] and total CO power spectra at $k=0.15\,$Mpc$^{-1}$ with $\Delta k=0.3\,$Mpc$^{-1}$ at 330\,GHz, against the masking volume for the different mock instruments. The [CII] is fixed at 0\% masking to minimise masking bias. The upper subplot shows the case using masks corresponding to the instrument $R$, with blue lines for $n=1$, orange lines for $n=2$, and solid/dashed/dotted lines denoting R\,=$100,\,200,\,300$. The lower subplot overlays the cases when using R\,$=100$ masks, using dark orange and blue for R\,$=200,\,300$. Masking at higher $n$ is increases the ratio by factor $2$. Higher R when using the standard mask sizes is as effective as R$\,=100$ but masks less volume. Using the R$\,=100$ masks results in improvement similar to changing from $n=1$ to $n=2$, though this is less pronounced for $n=2$, as that already successfully masks leakage.}
 \label{fig:DiffBeamImpact}
\end{figure}

These results are shown in Figure \ref{fig:DiffBeamImpact}. Increased $n$ improves the ratio of [CII]/CO by up to factor 2, indicating that masking is more effective at cleaning CO when their signal is more concentrated. Increasing R under the standard masking procedure does not clean more signal than R\,$=100$ (with approximately 0.25\,dex variance to the R\,$=100$ result), but achieves similar results with less volume coverage. This reduces the impact of volume normalisation on the power spectra, so the recovered results are more representative. The impact of this improvement is non-linear, as the change from R\,$=100$ to 200 is more pronounced than from 200 to 300. Higher R can still be used for cleaning though, due to the flexibility in how masks are applied: when using R\,$=100$ width masks, they are also effective at removing CO leakage. Compared to the base R\,$=100$ $n=1$ case there is an increase in ratio of factor 1.5, giving similar results to R\,$=100$ $n=2$. However, for $n=2$, there is minimal improvement as the wings of CO emission have been masked already. 

In general, we see the strongest increase in the [CII]/CO ratio for $n=2$, but higher R can give equivalent results with additional flexibility for lower masking volume. However, there are tradeoffs for this improvement. Such an instrument design requires increasing the total number of detectors per unit area, which is logistically difficult to implement whilst ensuring sufficient semiconductor cooling, and so without compromising sensitivity. In a similar manner, the current version of Prime-Cam shifts between 16 different FPI steps. With an increased number of detectors in frequency space, this would require more steps, or subdividing photons to a greater degree (depending on exact design), thereby requiring more time per area on the night sky. Consequently, to ensure the same sensitivity, the sky area coverage or total frequency range would need to be reduced in proportion with the increased R. Fewer of these issues would be the case for higher $n$, outside the difficulty of manufacturing said components with minimal leakage. In summary, while improving both R and $n$ would be ideal, with greater R providing more flexibility, higher $n$ would be ``safer'' for masking specifically (this may not reflect all LIM statistics).
\section{Supplemental Figure}\label{appendix: Extra Figures}

We show the full version of Fig. \ref{fig:ASPECSComparison} which was abridged in the main text.

\begin{figure*}[t]
 \centering
 \includegraphics[width=\linewidth]{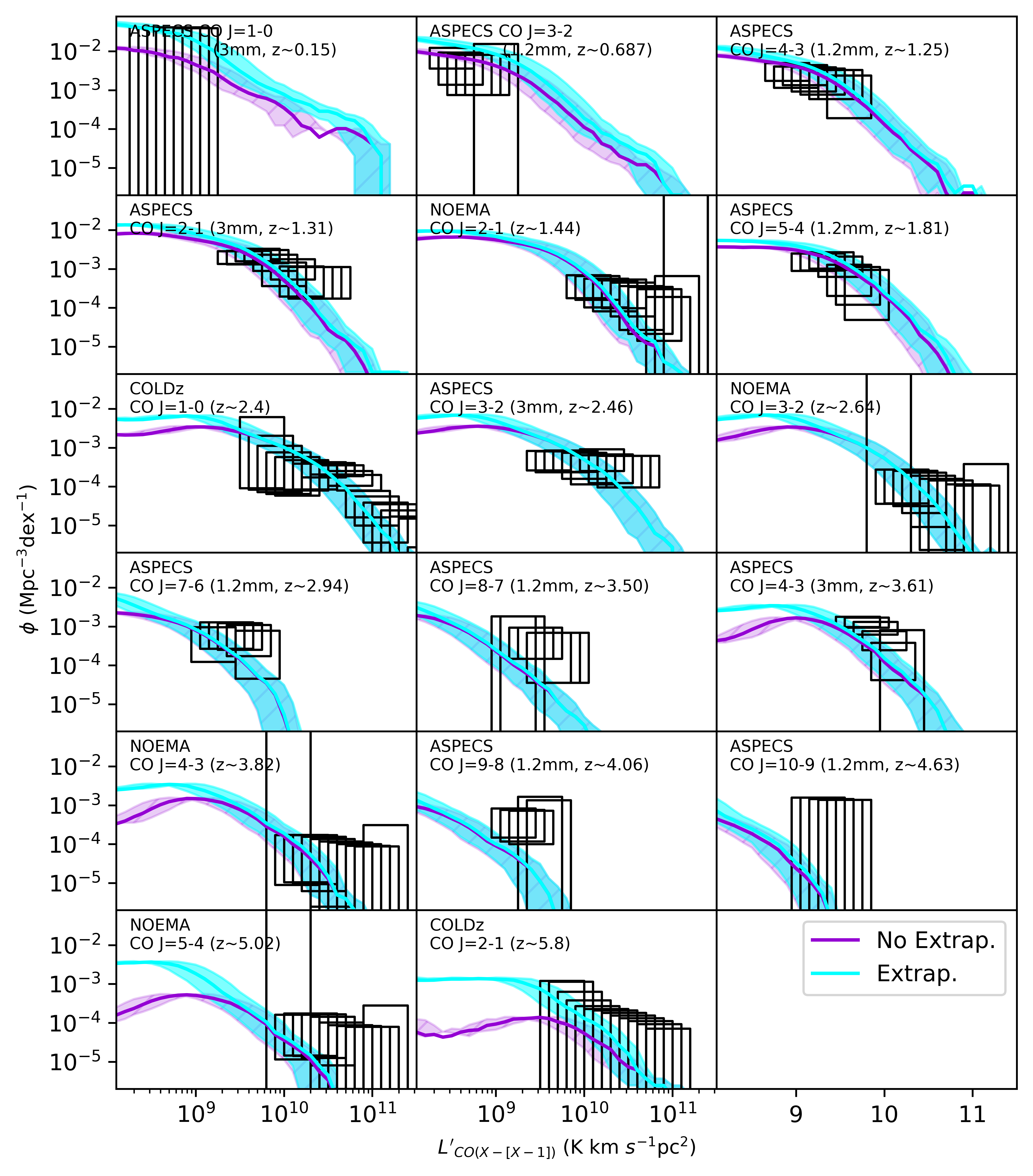}
 \captionof{figure}{Full version of Fig. \ref{fig:ASPECSComparison}.}
 \label{fig:ASPECSComparisonFULL}
\end{figure*}

\end{document}